\documentclass[preprint,authoryear,12pt]{elsarticle}
\usepackage{amsmath}
\begin{document}

\begin{frontmatter}

\title{Dynamics for Galactic Archaeology}

\author{James Binney}

\address{Rudolf Peierls Centre for Theoretical Physics, Keble Road, Oxford
OX1 3NP, England}

\begin{abstract}
Our Galaxy is a complex machine in which several processes operate
simultaneously: metal-poor gas is accreted, is chemically enriched by dying
stars, and then drifts inwards, surrendering its angular momentum to stars;
new stars are formed on nearly circular orbits in the equatorial plane and
then diffuse through orbit space to eccentric and inclined orbits; the
central stellar bar surrenders angular momentum to the surrounding disc and
dark halo while acquiring angular momentum from inspiralling gas; the outer
parts of the disc are constantly disturbed by satellite objects, both
luminous and dark, as they sweep through pericentre. We review the conceptual
tools required to bring these complex happenings into focus. Our first
concern must be the construction of equilibrium models of the Galaxy, for
upon these hang our hopes of determining the Galaxy's mean gravitational
field, which is required for every subsequent step. Ideally our equilibrium
model should be formulated so that the secular evolution of the system can be
modelled with perturbation theory. Such theory can be used to understand how
stars diffuse through orbit space from either the thin gas disc in which we
presume disc stars formed, or the debris of an accreted object, the presumed
origin of many halo stars.  Coupling this understanding to the still very
uncertain predictions of the theory of
stellar evolution and nucleosynthesis, we can finally extract a complete
model of the chemodynamic evolution of our reasonably generic Galaxy. We
discuss the relation of such a model to cosmological simulations of galaxy
formation, which provide general guidance but cannot be relied on for
quantitative detail.
\end{abstract}

\end{frontmatter}

\input binneymac
\section{Introduction}
\label{sec:intro}

In this article I will focus on the aspects of Galactic dynamics which seem
most relevant to Galactic archaeology. The latter is the endeavour to unravel
the Galaxy's history by measuring the positions, velocities, chemical
compositions and if possible the ages of stars. From these data we can in
principle determine the orbits of stars, so a major focus must be on how we
should characterise orbits and what we can infer about a star's history from
its orbit. The latter raises the questions ``how long has the star been on
this orbit'' and ``what was its original Galactic orbit?''.  It seems likely
that most, perhaps all, disc stars were formed on nearly circular orbits in
the plane. By contrast, many, perhaps most, halo stars were released onto
eccentric and inclined orbits on the tidal disruption by the Galaxy of the
system in which they were born. Therefore in addition to discussing how we
can determine the current orbits of stars, we must be able to compute the
probability that during the lifetime of a star that orbit can have evolved
from either an in-plane circular orbit or a particular eccentric inclined
orbit. 

For a variety of reasons, mostly unconnected with our Galaxy, we believe that
dark, unobserved matter plays a major role in generating the gravitational
field in which stars move. A major issue is the extent to which we can use
measurements of stars to constrain the spatial distribution of dark matter.
Another important question is the extent to which the distribution of dark
matter has been modified by gravitational interaction with gas and stars. 

The central assumption of Galactic dynamics is that the galaxy is
statistically in a steady state. In reality this assumption is false because
the central rotating bar, spiral structure in the disc and infalling objects
all cause the distribution of matter to evolve and thus the gravitational
field to be time-dependent. However, our strategy must be to start from the
assumption of a steady state and then to use perturbation theory to
understand how time-dependent effects modify a steady-state model.

This strategy is possible because the gravitational force is a long-range
one, so the force on a star is generally dominated by countless very distant stars
rather than by its near neighbours.\footnote{The force on a binary star is
dominated by its companion, but in this case we consider the motion of the
centre of mass of the binary rather than that of either component. Also
during a rare close encounter of field stars the forces on each star can be
dominated by that from its nearest neighbour, but such encounters are so
short-lived that long-term influence of the other star is small}
Consequently, we can model the Galaxy's gravitational field by the field that
would be generated by a system with the smooth density distribution $\rho(\vx)$
that would result from smearing the masses of stars and dark matter smoothly
through interstellar space. When it is possible to model the
gravitational field in this way, we say the system is \deffn{collisionless}.

\section{Orbits}\label{sec:orbits}

The simplest useful models of the Galaxy's gravitational field are
axisymmetric, so it is important to understand the nature of orbits in
axisymmetric gravitational potentials $\Phi(R,z)$. Since the component of
angular momentum $L_z$ about the potential's symmetry axis $\ve_z$ is
conserved, motion in an axisymmetric potential can be reduced to motion in
the $(R,z)$ plane. This is governed by the Hamiltonian
 \[
H=\fracj12(p_R^2+p_z^2)+\Phi_{\rm eff}(R,z)
\]
 where $p_R=\dot R$, $p_z=\dot z$ and the \deffn{effective potential} is
 \[
\Phi_{\rm eff}(R,z)=\Phi(R,z)+{L_z^2\over2 R^2}.
\]
 Numerical integration of orbits in typical effective potentials reveals that
the time series $R(t)$ and $z(t)$ are \deffn{quasiperiodic}, that is their
Fourier transforms
 \[
\widetilde{R}(\omega),\quad\widetilde{z}(\omega)
\]
 contain only discrete frequencies $\omega_i$. Moreover, these frequencies can be
expressed as integer linear combinations of two \deffn{fundamental
frequencies}. In the generic non-resonant case, the fundamental frequencies
can be taken to be the radial frequency $\Omega_r$, at which the star
oscillates in and out, and the vertical frequency $\Omega_z$ at which it
oscillated perpendicular to the potential's symmetry plane. Thus in the
generic case all the frequencies $\omega_i$ that occur in the Fourier transforms
of $R(t)$ and $z(t)$ can be written
 \[
\omega_i=n_r(i)\Omega_r+n_z(i)\Omega_z,
\]
 where $n_r(i)$ and $n_z(i)$ are both (possibly negative) integers.

When orbits in a Hamiltonian with $n$ degrees of freedom are quasi-periodic,
one may show \citep{Arnold} that there are $n$ integrals of motion -- smooth
functions on phase space $I_j(\vx,\vp)$ that are constant along each orbit:
$\d I_j/\d t=0$. Obviously one of these constants of motion can be taken to
be the Hamiltonian $H(\vx,\vp)$. However, any function of integrals is itself
an integral, so once we have more than one integral we have infinitely many
integrals to choose from. It proves expedient to choose integrals that can be
complemented by canonically conjugate variables, for if we use such
coordinates only half a star's coordinates evolve as the star moves because
the momenta are constants of motion. Integrals that can be complemented with
conjugate coordinates are called \deffn{action integrals} $\vJ$ and the
conjugate variables are called \deffn{angle variables} $\vtheta$. From
Hamilton's equations 
\[
\dot\vJ={\pa H\over\pa\vtheta}=0\quad;\quad
\dot\vtheta={\pa H\over\pa\vJ}=\vOmega(\vJ)\quad(\hbox{a constant}),
\]
 it follows that the angle variables increase linearly in time, the fundamental
frequencies playing the role of constants of proportionality. Thus, for
example
 \[\label{eq:thetaoft}
\theta_r(t)=\theta_r(0)+\Omega_rt\quad;\quad
\theta_z(t)=\theta_z(0)+\Omega_zt.
\]
 
Once we require integrals to be associated with conjugate variables and
position in phase space to be $2\pi$-periodic in the $\theta_i$, our choice
of integrals narrows dramatically. In fact, we are then confined to the
{action integrals} $J_r$ and $J_z$ that quantify the extent of the star's
radial and vertical oscillations, and integer linear combinations of these
basic actions. In practice the near uniqueness of action integrals is a
significant advantage.

It turns out that the two-dimensional surfaces $J_r=\hbox{constant}$,
$J_z=\hbox{constant}$ in the four-dimensional phase space $(R,p_R,z,p_z)$ can
be mapped 1:1 and differentiably onto a 2-torus (doughnut); that is, a
surface of constant $J_r,J_z$ has the topology of a 2-torus.  When (as is
generically the case) the fundamental frequencies are \deffn{incommensurable}
(so $\Omega_R/\Omega_z\ne n_R/n_z$ for any integers $n_R,n_z$), the star
eventually comes arbitrarily close to every point on the torus to which it is
confined by its integrals $J_r,J_z$.  Consequently, the orbit is essentially
identical with the torus, and we identify quasi-periodic orbits with
\deffn{orbital tori}.

Angle-action variables were introduced by Hamilton and Jacobi in the first
half of the 19th century as tools for the study of the dynamics of the solar
system. The underlying Hamiltonian of that problem is the Kepler Hamiltonian,
which governs the dynamics of a particle that moves in the gravitational
field of a point mass. For that Hamiltonian it is possible to express
ordinary polar coordinates $(r,p_r,\vartheta,p_\vartheta,\phi,p_\phi)$ as
analytic functions of $\vJ$ and $\vtheta$. Unfortunately, such formulae exist
for very few Hamiltonians -- essentially the $n$-dimensional harmonic
oscillator and the spherical isochrone Hamiltonian, of which the Kepler
Hamiltonian is a special case. None of these Hamiltonians is very close to
that of a flattened Galaxy, so we must manage without explicit formulae for
angle-action variables. Historically, this lack of formulae has discouraged
people from using angle-action coordinates but, as I will detail below, there
are now some effective work-arounds.

\subsection{Integrable systems and resonances}\label{sec:int_res}

A Hamiltonian that admits a global set of angle-action variables is said to
be \deffn{integrable}. The Kepler Hamiltonian is integrable, as are the
Hamiltonians associated with the gravitational potentials of certain
flattened or triaxial massive ellipsoidal bodies -- \cite{deZeeuw} discusses
these \deffn{St\"ackel potentials} in some detail. The Hamiltonians of real
galaxies are almost certainly not integrable, but they are best understood as
perturbations of integrable Hamiltonians, just as we consider the Hamiltonian
of the solar system to be the Kepler Hamiltonian defined by the Sun perturbed
by the gravitational fields of Jupiter, Saturn, etc., as well as relativistic
corrections to Newtonian mechanics.

The impact on an orbit of adding a perturbation may not be a smooth function
of the magnitude of the perturbation. This fact is of fundamental importance
for perturbation theory because it implies that the usual procedure for doing
perturbation theory -- expanding all variables as power series in the
``coupling constant'' that specifies the strength of the perturbation -- will in
general fail because the perturbed motion is in fact not analytic in the coupling
constant. 

The reason the response to a perturbation can depend discontinuously on the
magnitude of the perturbation is the quasi-periodic nature of the underlying
motion: we should think of the unperturbed motion as oscillatory at the
frequencies $\omega_i$ that occur in the Fourier decompositions of the
coordinates. If the perturbation is modulated at a frequency $\omega$ that
lies close to any of the $\omega_i$, the frequency $\omega-\omega_i$ at which
the orbiting particle perceives the perturbation can become very small.
Consequently, the particle can be pulled in the same sense for a long time
$\sim\pi/|\omega-\omega_i|$ and even a small perturbation can have a large
cumulative effect before it changes sign. In particular, for a critical
magnitude of the perturbation the motion can become \deffn{resonantly
trapped} by the perturbation, at which point the perturbation changes the
motion qualitatively and not merely quantitatively. The classic paradigm for
such trapping is the pendulum, which at a given energy will oscillate if
gravity is strong, but circulate if gravity is sufficiently weak. In
\S\ref{sec:ptheory} we explain the mathematical theory of resonant trapping.
If no resonant trapping occurs, the perturbation has simply transformed one
integrable Hamiltonian into another. Once resonant trapping has occurred, the
Hamiltonian is no longer integrable.

We can discover how important resonant trapping is for a given Hamiltonian,
and thus how close to integrable that Hamiltonian is, by numerically
integrating numbers of orbits and, if the motion is two-dimensional,
inspecting a \deffn{surface of section}, or if the motion is
three-dimensional, examining a \deffn{frequency map} -- these tools are
described in \S\S3.2.2 and 3.7.3(b) of \cite{BT08}.  Surfaces of section
reveal that realistic, axisymmetric models of the Galaxy's potential are
close to integrable. Surfaces of section for motion in planar rotating bars,
or frequency maps for motion in realistic models of the Galaxy's bar reveal
that in the vicinity of the bar's \deffn{corotation resonance} (where the
circular frequency in the axisymmetrised Galaxy model coincides with the
bar's pattern speed $\Omega_{\rm p}$) the Hamiltonian is far from
integrable.

An integrable Hamiltonian admits a global set of angle-action coordinates. 
Orbits that have been trapped by a particular resonance form an orbit family
that requires its own system of angle-action coordinates. Techniques for
constructing such coordinates were discussed by \cite{Kaasalainen95a}.

\subsection{Chaos}\label{sec:chaos}

When a single orbit in an integrable Hamiltonian is exposed simultaneously to
significant perturbations at more than one nearly resonant frequency
(``resonance overlap''), irregular behaviour can ensue \citep{Chirikov79}.
The standard interpretation of this \deffn{chaotic} behaviour is that the
orbit becomes trapped first by one resonance and then by another, with the
times of transition between resonantly trapped families varying erratically.
In classic cases one can show that numerically integrated orbits do shift
from one orbit family to another in an erratic way. Chaos is an important
phenomenon for Galaxy dynamics in the region just inside the corotation
resonance. By contrast chaos is unimportant in the immediate vicinity of the
Sun.

\subsection{What  integrable Hamiltonian?}\label{sec:intH}

The account of orbits we have given hinges crucially on the concepts of
resonance and perturbation amplitude. These concepts presuppose the existence
of an integrable Hamiltonian $H_0$, which endows orbits with fundamental
frequencies and defines the perturbing Hamiltonian as the difference $\delta
H=H-H_0$ between the actual and integrable Hamiltonians. 

In studies of planetary systems the Kepler Hamiltonian clearly stands out as
$H_0$, and in plasma and accelerator physics the harmonic-oscillator
Hamiltonian can plausibly serve as $H_0$. But for decades the integrable
Hamiltonian remained a shadowy, unidentified feature of Galactic dynamics. A
spherical Hamiltonian has been used \citep{Saha,Weinberg}, but this choice is
highly unsatisfactory for the Galaxy because it implies that $\Omega_z$ is
identically equal to the azimuthal frequency $\Omega_\phi$, whereas the
majority of Galactic orbits $\Omega_z>\Omega_\phi$, and this inequality
qualitatively changes the pattern of resonances. In studies of the solar
neighbourhood the harmonic oscillator has been used for $H_0$, but this
choice is unsatisfactory because it requires $\Omega_R$ and $\Omega_z$ to be
amplitude-independent whereas in the solar neighbourhood $\Omega_z$ declines
rapidly with amplitude.

A significant step forward was taken when \cite{deZeeuw} showed that
St\"ackel potentials could be generated by objects with mass distributions
that are similar to those of real galaxies, so St\"ackel Hamiltonians can be
used for $H_0$. One drawback of St\"ackel potentials is that they enforce a
relationship between the radial density profile and the ellipticity of the
isodensity surfaces of the generating body: the latter can be self-similar
only if asymptotically the density declines as $\rho\sim r^{-4}$; in real
galaxies we have $\rho\sim r^{-\alpha}$ with $3<\alpha<2$, and in this case
the isodensity surfaces of a body that generates a St\"ackel potential have
to become steadily more spherical as $r\to\infty$. Another drawback is that
for a St\"ackel Hamiltonian the transformations between angle-action and
spherical coordinates involve the numerical evaluation of one-dimensional
integrals, so St\"ackel Hamiltonians are not very convenient from a technical
perspective.

\cite{KaasalainenB94b} introduced a general technique for {\it constructing\/}
an integrable Hamiltonian that is close to any given galactic Hamiltonian.
\cite{Kaasalainen94} showed that on account of the closeness of the constructed
Hamiltonian to the given Hamiltonian, beautiful results for resonant orbits
and chaotic regions could be obtained from (suitably modified) first-order
perturbation theory.

Before explaining how integrable Hamiltonians can be constructed, we recall
some facts about coordinate transformations. The transformation
$(\vx,\vp)\leftrightarrow(\vtheta,\vJ)$ between ordinary phase-space
coordinates and angle-action coordinates is a \deffn{canonical
transformation}.\footnote{A coordinate system $(\vQ,\vP)$ is canonical iff the
Poisson bracket of two functions on phase space $f,g$ takes the form
 \[
[f,g]=\sum_i{\pa f\over\pa Q_i}{\pa g\over\pa P_i}-
{\pa f\over\pa P_i}{\pa g\over\pa Q_i}.\nonumber
\]
 A canonical transformation is simply one between canonical coordinate
systems.} Every canonical transformation can be obtained from a generating
function, and in our case we may choose to use a generating function of
the form $S(\vJ,\vx)$. It then generates the
transformation through
 \[
\vp={\pa S\over\pa\vx}\quad;\quad \vtheta={\pa S\over\pa\vJ}.
\]
  The classical way to find angle-action coordinates is to use the first of
these equations to eliminate $\vp$ from   $H(\vx,\vp)=E$ and to solve the
resulting \deffn{Hamilton-Jacobi} equation as a non-linear partial
differential equation for $S$. 

The Kaasalainen--Binney scheme for constructing an integrable Hamiltonian
starts from angle-action coordinates $(\vtheta',\vJ')$ obtained in this way
for some \deffn{toy Hamiltonian} $H_{\rm t}$, which may be the
isochrone Hamiltonian, or the harmonic oscillator Hamiltonian or a
St\"ackel Hamiltonian. Then a second generating function $S(\vJ,\vtheta')$ is
numerically constructed that maps the tori $\vJ'=\hbox{constant}$ of $H_{\rm
t}$ into the
tori $\vJ=\hbox{constant}$ of the actual or \deffn{target Hamiltonian}. The
numerically-constructed generating function has the form
 \[
S(\vJ,\vtheta')=\vtheta'\cdot\vJ+\sum_\vn s_\vn(\vJ)\e^{\i\vn\cdot\vtheta'}.
\]
 Here $\vn$ is a vector with integer components and the $s_\vn$, being
functions of the orbit's constant action $\vJ$, are numbers characteristic of
the orbit. The construction consists in solving for these numbers. The true
angle variables are
 \[
\vtheta={\pa S\over\pa\vJ}=\vtheta'
+\sum_\vn{\pa s_\vn\over\pa\vJ}\e^{\i\vn\cdot\vtheta'}.
\]
 Therefore we also have to solve for the derivatives of $s_\vn$, which we do
in a separate step \citep{BinneyKumar,KaasalainenB94a}. 

For any values of the
$s_\vn$ the mapping
 \[
(\vtheta,\vJ)\rightarrow (\vtheta',\vJ')\rightarrow(\vx,\vp)
\]
 maps the torus $\vJ=\hbox{constant}$ into ordinary phase space, and a
sufficient condition for determining the $s_\vn$ is to require that the
actual Hamiltonian $H(\vx,\vp)$ is constant on this torus. In practice the
Marquardt-Levenberg algorithm \citep{Pressetal} is used to minimise the
variance of $H$ over the image torus.  It is not possible with a finite set
of $s_\vn$ to achieve complete constancy of the Hamiltonian, but one can come
close to this goal.

By carrying out this fitting procedure at each point on a grid of actions
$\vJ$, we foliate phase space with tori on which $H$ is nearly constant.  In
principle further tori can be constructed by interpolating between the values
$s_\vn(\vJ)$ taken by the $s_\vn$ at the grid points. Thus we can completely
fill phase space with tori on which $H$ is approximately constant. The mean
value of $H$ on each of these tori defines an integrable Hamiltonian
$\overline{H}(\vJ)$ and the difference $\delta
H(\vtheta,\vJ)=H(\vtheta,\vJ)-\overline{H}(\vJ)$ is a well defined
perturbation.

\subsection{What technology to use?}\label{sec:whatTech}

Enough of general theory -- how should we proceed in practice to interpret
observational data in terms of angle-action variables? The first step is to
decide whether one wants to proceed from $(\vx,\vp)$ to $(\vtheta,\vJ)$ or in
the opposite direction. Numerically constructed tori provide the natural way
to go from $(\vtheta,\vJ)$ to $(\vx,\vp)$ because the output from the torus
machine is analytic expressions for $(\vx,\vp)$ given $(\vtheta,\vJ)$. For
example, a mock catalogue can be created from a model Galaxy by  choosing
actions $\vJ$ that sample the
distribution function $f(\vJ)$ (discussed below) and then
choosing the $\theta_i$ uniformly within $(0,2\pi)$. The numerically
constructed torus then yields $(\vx,\vp)$ and a star drawn from the
population described by the \df\ can be placed there. If you want to determine
the velocity distribution at some point $\vx$ in the model, for each sampled
$\vJ$ you can determine the values of $\vtheta$ (there are generally four or
six) at which the star will pass through $\vx$ and then compute the velocity
at which each passage will be made.

Sometimes one will want to pass from $(\vx,\vp)$ to $(\vtheta,\vJ)$. This can
be done with numerically constructed tori \citep[e.g.][]{McMillanB08} but
doing so can be slow because it involves an iterative search for the torus
that passes through the given $\vx$ with the given $\vp$. For disc stars
\cite{Binney10} used the ``adiabatic approximation'' to estimate $\vJ$ from
$(\vx,\vp)$ but this approach has now been superseded by a faster and more
accurate approximation based on St\"ackel potentials \citep{Binney12a}. A
related alternative is offered by \cite{Sanders12}, who explicitly fits a
St\"ackel potential to the Galactic potential in the region covered by the
given orbit.

These studies are restricted to axisymmetric models, which probably provide a
good basis for understanding the extended solar neighbourhood and outer
Galaxy, but will not deal with the Galaxy's inner few kiloparsecs, which are
dominated by the bar.  Tori have been numerically fitted to non-rotating
\citep{KaasalainenB94a} and rotating \citep{Kaasalainen95b} planar bars, but
the extension to three-dimensional rotating bars has yet to be made. In
principle this extension is straightforward.

\section{Equilibrium models}\label{sec:equimodels}

Models in which the Galaxy is in a perfectly steady state play a fundamental
role, not least because a good model of the Galaxy's potential $\Phi(\vx)$ is
fundamental to any investigation, and since $\Phi$ is substantially generated
by dark matter that we cannot directly observe,  we can determine $\Phi$ only
to the extent that  it is possible to argue
that the phase-space distribution of stars is statistically in equilibrium.

Let $f(\vx,\vp)\,\d^3\vx\d^3\vp$ be the probability of finding a star of some
particular type (specified by luminosity, colour, metalliticity, etc) in the
phase-space volume $\d^3\vx\d^3\vp$ at $(\vx,\vp)$. Then $f$ is called the
\deffn{distribution function} (\deffn{\df}) of that population of stars. By
Jeans' theorem \citep[][\S4.2]{BT08} a steady-state \df\ can depend only on
the actions, so $f=f(\vJ)$. Moreover, the Jacobian
$\pa(\vx,\vp)/\pa(\vtheta,\vJ)$ is unity and the angle variables all cover the
range $(0,2\pi)$ so the probability that a star of the given population lies
in the volume $\d^3\vJ$ of action space around $\vJ$ is
 \[
\d P(J)=\biggl(\int_\vtheta\d^3\vtheta\biggr)\d^3\vJ\,f(\vJ)
=(2\pi)^3\d^3\vJ\,f(\vJ).
\]
 Hence, $(2\pi)^3f(\vJ)$ is the probability density of stars of the given
population in three-dimensional action space. In other words, a steady-state
model consists of a particular distribution of stars of each type in
three-dimensional action space.  It is in these terms that we should think of
Galaxy models.

Conversely a steady-state model can be constructed by prescribing
distributions of stars in action space and then calculating the spatial
distribution of the stars and the kinematics that they have at every point.
Crucially, the spatial and kinematic properties of a
population are firmly tied to one another by the potential $\Phi$ (which
determines the structure of the orbital tori). 

\subsection{Disc DFs}\label{sec:discDF}

\cite{Binney10,Binney12b} showed that
data for the solar neighbourhood can  be largely accounted for by models
synthesised from \deffn{quasi-isothermal} \df s. Specifically, we write
 \[\label{eq:qi}
f(J_r,J_z,L_z)=f_{\sigma_r}(J_r,L_z)f_{\sigma_z}(J_z,L_z),
\] 
 where $f_{\sigma_r}$ and $f_{\sigma_z}$ are defined to be
 \[\label{planeDF}
f_{\sigma_r}(J_r,L_z)\equiv{\Omega\Sigma\over2\pi M\sigma_r^2\kappa}
[1+\tanh(L_z/L_0)]\e^{-\kappa J_r/\sigma_r^2}
\]
 and
 \[\label{basicvert}
f_{\sigma_z}(J_z,L_z)\equiv{\nu\over2\pi\sigma_z^2}\,\e^{-\nu J_z/\sigma_z^2}.
\]
 Here $\Omega(L_z)$, $\kappa(L_z)$ and $\nu(L_z)$ are, respectively, the
circular, radial and vertical epicycle frequencies of the circular orbit
with angular momentum $L_z$ and radius $\Rc(L_z)$, while
\[\label{eq:defsSigma}
\Sigma(L_z)=\Sigma_0\e^{-\Rc/R_\d}
\]
 governs the approximately exponential surface density of the disc and the
disc's mass is
 \[
M=2\pi\Sigma_0 R_\d^2.
\]
 The
functions $\sigma_r(L_z)$ and $\sigma_z(L_z)$ control the radial and vertical
velocity dispersions in the disc and are approximately equal to them at
$\Rc$. Given that the scale heights of galactic discs do not vary strongly
with radius \citep{vdKSearle}, these quantities must increase inwards. The
natural Ansatz to achieve this is
 \[
\sigma_r(L_z)=\sigma_{r0}\,\e^{(R_0-\Rc)/R_{\sigma r}}\quad;\quad
\sigma_z(L_z)=\sigma_{z0}\,\e^{(R_0-\Rc)/R_{\sigma z}},
\]
 which imply that the velocity dispersions decline outwards exponentially,
with scale length $R_{\sigma i}$. Our expectation is that $R_{\sigma
i}\sim2R_\d$ and in the simplest models one assumes that $R_{\sigma
r}=R_{\sigma z}$.

The functions $f_{\sigma_i}$ satisfy the normalisation conditions
 \[
\int_0^\infty\d J_r\,f_{\sigma_r}
={\Omega\Sigma\over2\pi M\kappa^2}[1+\tanh(L_z/L_0)]\quad;\quad
\int_0^\infty\d J_z\,f_{\sigma_z}={1\over2\pi}.
\] 
 Consequently, the number of stars per unit angular momentum
\[
g(L_z)\equiv\int\d J_r\int\d J_z\,f(J_r,J_z,L_z),
\]
 decreases as
$\sim\Sigma(L_z)/\kappa(L_z)$, so roughly exponentially. Fig.~1 of
\cite{Binney12b} shows that in a realistic potential $\Phi(R,z)$ a
quasi-isothermal \df\ with $R_\d/R_\sigma=0.45$ generates a disc in which, to
a good approximation, the density declines exponentially with both $R$ and
$|z|$ and the vertical velocity dispersion $\sigma_z$ is essentially
independent of $z$ for $|z|\lta600\pc$ and rises very slowly at greater
distances from the plane. Hence a quasi-isothermal \df\ provides a rigorous
mathematical realisation of the isothermal mono-abundance components into
which \cite{Bovy12a} argue the disc should be decomposed as an alternative to
the traditional divide into thin and thick discs.

\cite{Binney10,Binney12b} represented the thick disc with a quasi-isothermal and the
thin disc as a superposition of quasi-isothermals, one for the stars of each
age $\tau$. From the analysis of Hipparcos stars by \cite{AumerB} he took the
age dependence of the velocity-dispersion parameter of a coeval cohort's
quasi-isothermal, and then had
 \begin{eqnarray}\label{sigofLtau}
\sigma_r(L_z,\tau)&=&\sigma_{r0}\left({\tau+\tau_1\over\tau_{\rm
m}+\tau_1}\right)^\beta\e^{(R_0-\Rc)/R_\sigma}\nonumber\\
\sigma_z(L_z,\tau)
&=&\sigma_{z0}\left({\tau+\tau_1\over\tau_{\rm m}+\tau_1}\right)^\beta
\e^{(R_0-\Rc)/R_\sigma}.
\end{eqnarray}
 Here $\sigma_{z0}$ is the approximate vertical velocity dispersion of local
stars at age $\tau_{\rm m}\simeq10\Gyr$, $\tau_1\simeq10\Myr$ sets velocity
dispersion at birth, and $\beta\simeq0.33$ is an index that determines how
the velocity dispersions grow with age.  He further assumed
that the star-formation rate in the thin disc has decreased exponentially
with time, with characteristic timescale $t_0$, so the thin-disc \df\ is
 \[\label{thinDF}
f_{\rm thn}(J_r,J_z,L_z)={\int_0^{\tau_{\rm m}}\d\tau\,\e^{\tau/t_0}
f_{\sigma_r}(J_r,L_z)f_{\sigma_z}(J_z,L_z)
\over t_0(\e^{\tau_{\rm m}/t_0}-1)},
\]
 where $\sigma_r$ and $\sigma_z$ depend on $L_z$ and $\tau$ through equation
(\ref{sigofLtau}).  The normalising constant $\Sigma_0$ that appears in
equation (\ref{eq:defsSigma}) he took to be the same for both discs and used
for the complete \df
 \[
f(J_r,J_z,L_z)=(1-\lambda)f_{\rm thn}(J_r,J_z,L_z)+\lambda f_{\rm thk}(J_r,J_z,L_z),
\]
 where $\lambda$ is a parameter that controls the fraction of stars that belong to
the thick disc.

In \cite{Binney12b} the values of the three parameters $\tau_1$, $\tau_{\rm
m}$ and $\beta$ in the thin-disc \df\ were set from \cite{AumerB} and the
remaining nine parameters in the overall \df, $\sigma_{r0}$, $\sigma_{z0}$,
$R_\d$ and $R_\sigma$ for each disc plus the thick-disc weighting $\lambda$, were
fitted to subsets of data for the solar cylinder using a plausible potential,
generated by stellar and gas discs, a bulge and a dark halo that begins to
dominate the radial force just inside $R_0$. The principal results from this
fitting exercise were:

\begin{itemize}
\item When the thick disc is omitted ($\lambda=0$) and the thin-disc \df\ is fitted
to the $U$, $V$ and $W$ distributions of stars in the Geneva-Copenhagen
Survey \citep[][hereafter GCS]{Nordstrom04,HolmbergNA}, the density profile $\rho(z)$ of the
thin disc is in good agreement with that derived for $|z|\lta500\pc$ from
star counts. This finding provides support for the adopted potential. 

\item The $U$ and $V$ distributions are fitted by the thin-disc \df\ as well
as can be expected given the presence in the solar neighbourhood of star
streams that clearly lie beyond the scope of an axisymmetric equilibrium
model. The wings of the observed $W$ distribution are under-populated by the
\df.

\item When a thick-disc component is added and used to improve the fits to
the GCS velocity distributions while at the same time fitting the density
profile $\rho(z)$ of \cite{GilmoreR}, the velocity and density constraints can
be fitted simultaneously to good precision. Moreover, the resulting model
does an excellent job {\it predicting\/} a preliminary estimate of
$\sigma_z(z)$ extracted from the RAdial Velocity Experiment
\citep[][hereafter RAVE]{Steinmetz} by \cite{Burnett}. The model also does a
reasonable job of predicting the distribution of $V$ components at
$|z|\simeq1\kpc$ published by \cite{Ivezic08}.

\item Since the wings of the GCS distributions of $U$ and $V$ had already
been filled out by thin disc, the fitted value of the thick disc's $\sigma_{r0}$
parameter ($26\kms$) was much smaller than the fitted value of $\sigma_{z0}$
($45\kms$). That is, the thick disc that led to successful predictions was
radially cooler than the thin disc (which had $\sigma_{r0}=40\kms$).  To find
a radially hot thick disc, the thin disc \df\ was set to values that did not
fill out the wings of GCS $U,V$ distributions and then the thick disc was
fitted as before. This resulted in $\sigma_{r0}$ being $30\kms$ for the thin
disc and $40\kms$ for the thick disc. This model predicted values of
$\sigma_\phi$ and $\ex{v}_\phi$ at $|z|\simeq1\kpc$ that were, respectively,
too large and too small. That is, a radially warm thick disc conflicts with
the data by requiring too little rotation and too much random motion at large
$|z|$.

\end{itemize}

Dividing the disc into components such as ``thick'' and ``thin'' only makes
sense in so far as one can track the phase-space distributions of
intrinsically distinguishable stars \citep[i.e., stars that may be distinguished by
\hbox{[Fe/H]} or \hbox{[$\alpha$/Fe]}, or age -- see][\S10.4.3]{BinneyM98}.
Since \cite{Binney12b} did not use data that distinguished stars by
metallicity or age, the principal interest of his exercise is methodological
-- it shows how one can model data in a dynamically consistent way and derive
testable predictions. Real progress will be made when the models are extended
to include different  \df s for several distinguishable groups of stars, such
as stars younger than $1\Gyr$, or with [Fe/H] in different ranges, etc.
Setting up such models is straightforward; the complex step is fitting them
to the data, for more sophisticated data are associated with strong selection
effects.

\subsection{DFs for the halos}\label{sec:haloDF}

In principle we should have \df s for both the stellar halo and the
dark-matter halo. The only dynamically consistent halo \df s in the
literature are functions $f(E,L_z)$ of energy and possibly azimuthal angular
momentum. Such a \df\ implies that $\sigma_R=\sigma_z$ everywhere, which
certainly does not hold for some classical denizens of the stellar halo, for
example RR-Lyrae stars \citep{Delhaye}. A partial solution to this problem
is to pretend the Galaxy's potential is spherical, so the total angular
momentum $L$ can be added to the list of isolating integrals
\citep[e.g.][]{DeasonBE,Gomez10}.  \cite{DeasonBE} have modelled the
kinematics of blue horizontal-branch (BHB) stars using the \df
 \[
f=[1+\chi\tanh(L_z/L_0)]L^{-2\beta}E^s,
\]
 where $\chi$, $L_0$, $\beta$ and $s$ are all parameters. While a \df\ of
this form permits the halo to display both systematic rotation and either
radial or tangential bias in the velocity-dispersion tensor, it is
implausible
in key respects. The power-law dependence on $L$ causes the phase-space
density to diverge as $L\to0$ in the radially biased case ($\beta>0$) and
gives rise to a bimodal distribution of tangential velocities in the case of
tangential bias.  On account of these pathologies the \df\ cannot be expected
to provide a good fit to observed velocity distributions, and when a model
does not provide a good fit, the physical significance of the best-fitting
parameters is doubtful.

Although useful qualitative insights into the halo's dynamics can be obtained
by approximating the Galaxy's potential as spherical
\citep[e.g.][]{BinneyPetrou}, when fitting real data we should recognise that
the Galaxy's potential is not spherical, and indeed that there is much
interest in determining its shape, and thus diagnosing the shape of the dark
halo. In a non-spherical potential at least one of the integrals that appears
in the \df\ has to be non-classical, and may as well be taken to be $J_z$. If
the potential is non-axisymmetric, we must use $J_\phi$ in place of its
limiting form for an axisymmetric system, $L_z$. Consequently, the only real
choice we need to make is whether to use as the third integral $E$ or $J_r$.
Using $J_r$ rather than $E$ has the following advantages
 \begin{itemize}
 \item In any potential the range of $J_r$ for bound orbits is $(0,\infty)$
whereas the range of $E$ is $(\Phi(0),0)$ and thus depends on the depth of
the adopted potential.  Consequently it is impossible to make general
statements about the physical implications a given \df\ $f(E,\ldots)$ in the
way that one can about a \df\ $f(\vJ)$.

\item Whereas $f(\vJ)$ is the density of stars in action space, $f(E,\ldots)$
is not the density of stars in $(E,\ldots)$ space because
$\d^3\vx\d^2\vp=\d^3\vtheta\d J_\phi\d J_z\d E/\Omega_r$.

\item Ultimately the Galaxy's potential must be made consistent with the
total density
 \[
\rho(\vx)=\sum_{{\rm pops}\ i}\vM_i\int\d^3\vp\,f_i,
\]
 where $\vM_i$ is the mass of the $i$th population and the populations summed
over include the dark matter. This requirement for self-consistency is easy
to satisfy when we adopt $f_i(\vJ)$: we take a trial potential, evaluate
$\rho$ in this potential, solve for the corresponding potential, and
re-determine $\rho$. This cycle converges after a handful of iterations. When
$f_i(E,\ldots)$ is given, considerable cunning is required to devise a convergent
scheme of this type \citep[see][\S4.4.2(b) for more detail]{Rowley,BT08}.

\item $\vJ$ is an adiabatic invariant whereas $E$ is not. Consequently, when
the potential changes slowly, through the growth of the stellar disc, for
example, $f(\vJ)$ is constant while $f(E,\ldots)$ is not. For this reason the
classic work of Spitzer's school on globular clusters \citep{Spitzer_globs}
used $J_r$ (usually denoted $q$) in numerical computations of cluster
evolution even though the original \df\ was taken to be $f(E,\ldots)$.
\end{itemize}

\cite{Pontzen} estimated the actions of particles in some large simulations
of cosmological dark-matter clustering using a spherically symmetrised
potential and reached the interesting conclusion that in the portion of
action space in which periods are less than the Hubble time, the \df\ is
simply a product of exponentials of the actions. Consequently, it seems that
there is a natural choice of functional form for the \df\ of the dark matter.

\section{Cosmological simulations}\label{sec:simulations}

As soon as electronic computers became widely available in the 1960s, they
had a big impact on galactic dynamics. \cite{HenonH} showed that
quasiperiodic and chaotic motion can exist side by side, while
\cite{HockneyH} showed that cool, self-gravitating  discs
develop strong bars on essentially a dynamical timescale. A decade later
N-body models showed that velocity anisotropy can flatten galaxies in the
absence of rotation and that triaxial stellar systems are dynamically
possible \citep{Binney76,AarsethB}.

The structure of an N-body model is implicit in the initial conditions from
which it is started, but these are typically far removed from the model's
final form, and the map from initial conditions to final equilibrium is not
easily understood. In these circumstances a natural way to proceed is to
imagine how a galaxy might have formed and to let this speculation suggest
suitable initial conditions. For example \cite{Gott} used initial conditions
associated with a ``top hat'' over-density in the expanding Universe, while
\cite{Binney76} adopted initial conditions inspired by Zeldovich's picture of
collapse to a pancake \citep{Zeldovich}.

 The advent of the Cold Dark Matter cosmology in the 1980s put the choice of
initial conditions for the dominant dark-matter component on a rigorous
footing, and the next twenty years were taken up using ever more
sophisticated software and faster hardware to understand the dynamical
evolution of dark matter (DM) from the given initial conditions. Consequently, we
now know what would happen in a universe free of baryons.

Unfortunately, the CDM cosmology does not make clean predictions for the
initial conditions of galaxies' stellar components because we are unable to
compute with any confidence where and at what velocities gas turns into
stars. The difficulty is that so long as matter remains gaseous it responds
to pressure forces as well as gravity, and pressure depends on the
temperature and density of gas. Hence to compute gas dynamics one needs to
follow the heating and cooling of gas. The gas is heated by adiabatic
compression and by shocks in which gas suffers an abrupt change of velocity.
The velocities driving shocks may be gravitationally driven but they are often
driven by outflows powered by supernovae and accreting objects. There is
abundant evidence that outflows have a big impact on galaxy formation, but
simulating them has proved extremely hard and can currently not be achieved
with standard physics: some kind of ad-hoc suspension of the standard
equations  is required to generate significant outflows, such as shutting
off cooling for a time or distributing by hand the energy of a supernovae through a
kiloparsec-sized volume.

Ab initio simulations of galaxy formation are exceedingly hard because on the
one hand the structure of the inner several kiloparsecs of the galaxy is
strongly influenced by material that is brought to it by flows that have a
characteristic length scale of megaparsecs, so to simulate the formation of a
galaxy one has to follow the dynamics of a section of the Universe at least
$10\Mpc$ across. On the other hand, to get the heating and cooling of gas
right it is essential to resolve shocks and contact discontinuities that have
sub-parsec scales. In short, simulations with enormous dynamic range are
required. Nobody knows what the minimum dynamics range is, but it is
certainly beyond the reach of current supercomputers.

Successive generations of stars synthesise heavy elements and the outflows
driven by star formation drive a portion of these elements away from the
sites of their synthesis. Heavy elements in the photospheres of low-mass,
long-lived stars provide a fossil record that could tell us much
about galaxy formation. Early attempts to understand the chemical evolution
of galaxies modelled the annulus of gas within the Galaxy with $R\sim R_0$ as
a box that initially contained just pristine gas but gradually contained more
and more stars and more highly enriched gas
\citep{vdBergh62,Schmidt63,Tinsley68,PagelP75}. An
early conclusion of these studies was that metal-poor gas continued to enter
the box throughout the history of the Universe. We return to this crucial
insight below.

Huge efforts have been made to understand the relative proportions of various
nuclides each generation of stars produced \citep{Thielemann,Nomoto} -- for a
single star these proportions depend on the star's initial mass mass (which
determines the nature of its death), its initial chemical composition (which
determines the extent to which nuclides heavier than Fe can be built up by
neutron capture), and potentially its speed of rotation (which affects
convection within the star). Decades of work on this problem have made it clear
that it is incredibly hard: not only does one require complete knowledge of
the evolution of interacting binary stars, but one needs to be able to
predict which parts of an exploding star will be ejected and which parts will
fall back to form the remnant, and nothing about supernovae is
straightforward -- they are distinctly non-spherical, turbulent relativistic,
and potentially highly magnetised non-equilibrium objects.  

Given that it is so hard to predict the mix of heavy elements that a given
generation of stars injects into the interstellar medium, a natural strategy
is to determine this mix empirically by observing the chemistry of supernova
remnants. This too is an extremely challenging programme because there are
many different kinds of supernovae, each of which has to be studied
independently from an appropriate supernova remnant, and it is uncertain
which kind of supernova created each of the observable remnants. Moreover,
the observationally accessible remnants probably do not cover all relevant
supernova types.

Two islands of clarity emerge from this sea of difficulty and uncertainty.
One is that deflagration supernovae, in which a C/O white dwarf suddenly
burns to iron-peak elements, (i) are major contributors to the cosmic Fe
production, and (ii) do not occur in large numbers until $\sim1\Gyr$ after a
population of stars is formed because time is required for a star of initial
mass $<8\msun$ to evolve to a white dwarf and further time is required for
its companion to dump significant mass on the white dwarf \citep{Podsi}.
Hence stars formed in the first $\sim\Gyr$ of a galaxy's life are Fe-poor
(which is perversely usually designated $\alpha$-rich).

The other island of clarity is that stars form in clusters and the
stars in a given cluster have a characteristic ``fingerprint'' of relative
abundances that reflects the particular mix of massive stars that enriched
the gas from which they formed \citep{FreemanJBH}. This fact suggests that we
could discover how stars move through phase space by dividing stars with
measured phase-space coordinates into groups with characteristic relative
abundances that probably formed in the same star-forming event.

We have seen that simulations of galaxy formation are formidably difficult,
even when no attempt is made to follow chemistry, and we have seen how far the
current theory of stellar evolution is from being able to predict reliably
the chemical evolution of the interstellar medium. Given that to model
current data one must {\it simultaneously} follow the dynamics and the
chemistry of the interstellar medium \citep[e.g.][]{Brook12}, it is clear that
definitive simulations of galaxy formation will not be available for many
years, perhaps never.  Nevertheless in the recent years significant progress
has been made with simulations of galaxy formation and we now have a pretty
good idea of how galaxies form \citep{MoWhite}. 

Dark matter leads the way by
collapsing on relatively small scales to form large numbers of cuspy dark
halos. These tumble together to form larger halos. When the ratio of the
masses of two halos is less than about 10, they merge rather completely
through a combination of tidal disruption and dynamical friction. But when a
large halo accretes much smaller halos, the latter are merely stripped of
their outer layers and they live on as reduced systems orbiting through the
outer halo of their massive host.

Gas falls into whatever dark halos are about, and loses energy by radiation.
So it tends to accumulate in the inner regions of dark halos. Once gas
dominates the local gravitational field, gravitational collapse driven by
radiative cooling runs away at various locations and clusters of stars are
formed. The most massive stars soon produce powerful outflows, which drive
away most of the gas from which they formed. Some of this gas will fall into
some other halo, other gas will return to the halo from it was ejected after
a considerable delay. Thus the tumbling together of dark halos is punctuated
by episodes of star formation that terminate when energy released by young
stars temporarily drives away residual gas.

In any region of space there are a few massive halos and many much smaller
ones, many of them orbiting within the massive halo. These orbiting halos
have little opportunity to pick up gas from which to form stars because the
ambient gas is moving in the dominant gravitational field of their host halo,
and their own gravitational field is too weak to grab such gas at the
relative velocity imposed by the host system. So the star-forming days of a
halo come to an end soon after it starts to orbit within a much more massive
halo.

Luminous galaxies are systems that have been the local centre of
gravitational attraction for a long time in a region of high ambient gas
density. In what were initially very over-dense regions, halos that are now
part of a rich cluster of galaxies early on found themselves in collapsing
regions of exceptional gas density. Consequently they accreted gas and formed
stars fast. In these regions the scale of halo clustering exceeded the scale
of an individual galaxy rather early on, and from this time even massive
galaxies have been orbiting in cluster-sized halos so they have been unable
to accrete gas and their star formation long ago faded. Hence in clusters we
find luminous galaxies that are red and dead. 

In lower density environments the scale of clustering increased more slowly
and gas was accreted more slowly, so here we find luminous galaxies that are
either still forming stars or only recently ceased to do so.

The atomic and molecular gas from which stars form has a natural velocity
scale $7-10\kms$: outflows from stars can sustain turbulence with this kind
of velocity dispersion. If the cool gas in a halo forms a rotating disc,
this disc will be thin if its orbital velocity is much greater than $10\kms$,
otherwise it will be thick and lumpy. Hence spiral galaxies are associated
with halos that have circular velocities in excess of $\sim80\kms$.

Ours is just such a galaxy. The visible Galaxy is the baryon-dominated core
of a much larger halo within which orbit many much smaller halos. There are
halos on orbits with periods much shorter than a Hubble time that have been
through pericentre several times, and ones that are falling in for the first
time. We expect halos to arrive in groups, and recent measurements of the
proper motion of the Magellanic Clouds \citep{Kallivayalil} suggest that the
Clouds form just such an infalling group. The Sgr Dwarf, by contrast, fell in
long ago and has already orbited the Galaxy several times. 

As halos orbit through the Galaxy tides pull out of them leading and trailing
streams, so the Galaxy's halo is tracery of such streams. Since the outer
parts of halos are DM-dominated, most stripped material is dark and stars are
likely to enter the streams only after the halo has been orbiting for some
time and has been stripped down to its core.

Several lines of argument imply that the outer Galaxy is filled with a
\deffn{corona} of plasma at the virial temperature $\sim2\times10^6\K$.
Although it is too tenuous to be detected directly through its X-ray emission
(in contrast to the analogous virial-temperature gas found in rich groups and
clusters of galaxies), the corona manifests itself by confining denser cold
gas that we can detect spectroscopically \citep{Sembach}, and by stripping
cold gas from the LMC/SMS system and forming it into a stream that stretches
right across the southern Galactic hemisphere. Two arguments indicate that
the Magellanic Stream is not a regular tidal stream: (i) it is not associated
with stars, presumably because any stars that were gravitationally stripped
alongside the gas were not subject to pressure forces and are consequently to
be found at some distance from the gas; (ii) the leading arm of the
Magellanic Stream is extremely short and does not in the least resemble the
trailing arm, as it would if it were a stellar stream and generated by
gravity alone.

The corona interacts with the gas disc in two ways: (i) hot plasma blown out
of the star-forming disc must join the corona and enrich it with freshly
synthesised heavy elements; (ii) clouds of cool ($\sim10^4\K$) gas that are
thrown $1-2\kpc$ off the disc by supernova bubbles interact strongly with
the corona as they move at velocities $\sim100\kms$ through the coronal gas.
In the first instance the flow of coronal plasma over the leading surfaces of
clouds ablates the clouds. Downstream of the clouds cool ablated gas from the
clouds mixes with coronal plasma in a wake. Depending on the density of the
plasma, mixing in the wake may eliminate all $\sim10^4\K$ gas. In this case
the density of the corona steadily increases. Eventually the density will
reach the regime in which the wake's ablated gas cools the coronal gas and there is
more gas at $\sim10^4\K$ in the wake than has been stripped from the cloud.
After $\sim100\Myr$ the residual cloud and its bloated wake fall back to the
disc and the net effect of the original ejection of the cloud has been to
augment the supply of cool star-forming gas in the disc \citep{Fraternali}.

Given the limitations of current simulations with respect to their inadequate
resolution and uncertain foundations in stellar evolution, 
the picture we have just drawn is necessarily qualitative rather than
quantitative. Moreover, an essential aspect of the picture is the stochastic
nature of the formation process that is driven by infall of objects great and
small. Hence we cannot even dream of one day having a precise simulation of
the formation of the MW; the most we can hope for is an ensemble of
simulations of which the MW might be a member, from a statistical
perspective. That is, we have to focus on the {\it statistics} of simulations
and develop tests that enable us to say that the MW is plausibly a member of
this set of simulations even though it differs from all the set's members.

What are he relevant statistics? Two might be the luminosity and mass
functions of the Galaxy's satellites. Another might be the amplitude and
pitch angle of the spiral arms, another the mean and rms rates of gas
accretion as functions of radius and time, and so on. In principle it should
be straightforward to determine such numbers from simulations, although huge
computational resources will be required to determine these numbers from
statistically significant samples of state-of-the-art simulations. Extracting
such numbers from observations of the MW is by no means straightforward and
will involve models that are refinements of the models we introduced in
\S\ref{sec:equimodels}

\section{Fitting models to data}\label{sec:datafitting}

Astronomy is probably the science that is most affected by selection effects,
and these effects are nowhere more important than in Galactic archaeology. We
observe from a particular point in the plane, and which objects enter
our catalogues is determined primarily by the ease of observation.
Consequently, a major effort is required to extract from the catalogues the
underlying frequencies of objects.

We inevitably select objects by spatial location -- magnitude selection
cannot be avoided, and the relationship between luminosity and apparent
magnitude depends on distance and extinction, and is consequently a strong
function of position. At a given location we can avoid selection by velocity,
but often we do not because it is hard to resist the temptation to sift
interesting targets from the generality by selecting on proper motion
\citep[e.g.][]{Sayres12}. 

The traditional way of handling selection effects is correction of the data.
For example, when determining the density profile $\rho(z)$ we multiply the
number of stars in the catalogue at $z$ by the inverse of the fraction of the
population's luminosity function that are bright enough to pass the survey's
magnitude limit. Such corrections always rely on prior information -- in this
case the population's luminosity function -- and the required information is
often precisely what we still need to determine. For example, spectroscopic
surveys of the local disc tend to select targets by proper motion
\citep{Sayres12} or space velocity \citep{Bensby}, and to correct these
data you need to know the complete velocity distribution of the underlying
population. Consequently, you need a model of the population. Hence Galaxy
models are absolutely fundamental to progress in Galactic archaeology. 

The models traditionally used in the interpretation of spectroscopic surveys
\citep[e.g.][]{Bensby,Bovy12b} are kinematic -- they specify the velocity
distribution without regard to Jeans theorem -- and such models are open to
two objections: (i) they have more degrees of freedom than they should, and
(ii) they are liable to specify physically unattainable velocity
distributions. An excessive number of degrees of freedom is both a nuisance
in that it decreases the strength of the inferences that can be drawn from a
given survey and a huge missed opportunity for the following reason. If you
measure the profiles $\rho(z)$ and $\sigma_z(z)$ of some population, you can
probably find a potential $\Phi(z)$ for which they are reasonably
consistent. If you now measure $\rho(z)$ or $\sigma_z(z)$ for any other
population, one of these functions will immediately lead to a testable
prediction of the other. Consequently, in a set of measured profiles
$\rho(z)$, $\sigma_z(z)$ for a number of populations, there is considerable
redundancy, and from the whole set it should be possible to determine
$\Phi(z)$ with good precision even in the presence of significant
observational errors in each data set. It follows that you cannot
consistently hypothesise both $\rho(z)$ and $\sigma_z(z)$ for more than one
population simultaneously.

Traditionally the positions and velocities of individual stars have been
inferred from the data, which invariably consist of photometry, astrometry
and sometimes spectroscopy. Then the stars are binned, either in phase space
or some projection of it such as real space, and the resulting stellar
densities are either compared with the predictions of models or analysed from
the perspective of Jeans' equations.  The data from the Sloan Digital Sky
Survey (SDSS) have been analysed in this way. \cite{Juric} showed that the
spatial density of stars is very similar to that expected from a
superposition of a thin and a thick disc. \cite{Ivezic08} analysed the
kinematics of these stars as a function of metallicity and showed that the
sample divides naturally into a metal-poor halo and a metal-richer disc with
distinct kinematics. \cite{Bovy12a} determined the spatial distribution of
stars as a function of their position in the $(\feh,\afe)$ plane and
concluded that each such population is consistent with forming a
double-exponential disc. \cite{Bovy12b} went on to argue that each such
mono-abundance component is an isothermal in the sense that within it
$\langle v_z^2\rangle$ is almost independent of height.

There is, however, a strong case for not attempting to reconstruct the
phase-space distribution directly from the data. The errors in the data
are always appreciable for the majority of stars in any catalogue because
most stars inevitably lie close to the catalogue's magnitude limit, which
will be close to the point at which measurements become too uncertain to be
useful. Non-negligible errors need to be properly accounted for when making
deductions from the data, and this is hard to do rigorously when directly
inverting the data.

The distance $s$ to a star is a single number that affects not only the
star's presumed location, but its presumed tangential velocity, luminosity,
reddening and therefore intrinsic colours also. Distances are always
significantly uncertain, and the error in $s$ induces correlated errors in
all these variables.  A clean error analysis in the presence of correlated
errors is hard. Still more worrying, the Gaia Catalogue will provide parallaxes
$\varpi$ for a billion stars, and many of these parallaxes will be negative.
A negative parallax does not yield a meaningful distance, but it does
constrain $s$: to $\gta1/\sigma_\varpi$. Moreover, many positive parallaxes
will not be much larger than their errors, and in this case the probability
distribution of $s$ is very skew, with a long tail to large $s$. We can
overcome all these difficulties by comparing the data to the predictions of
models in the space of observables
$(\alpha,\delta,\varpi,v_\parallel,\mu_\alpha,\mu_\delta,V,B-V,\ldots)$, for here
the errors will be independent and, by the central limit theorem, to a good
approximation Gaussian.

The downside of working in the space of observables is that physically
significant information becomes non-local -- an over-density of stars at some
location in real space maps into less pronounced enhancements in the numbers
of stars over a wide range of apparent magnitudes.

\cite{McMillanB12} used orbital tori to investigate our ability to infer the
\df\ of a population from various combinations of photometry, astrometry and
measurements of line-of-sight velocities. They assumed throughout that the
potential is known and that the stars are drawn from a very broad luminosity
function, so the apparent magnitude of a star conveys very little information
about $s$ -- the photometry of the SDSS constrains distances much more
strongly, for example. Notwithstanding the breadth of their assumed
luminosity function, \cite{McMillanB12} found that photometry and proper
motions with Gaia-like errors for $10\,000$ stars widely distributed over the
sky are sufficient to constrain the \df\ of the population very narrowly.
Adding parallaxes and line-of-sight velocities merely diminishes the already
modest errors on the parameters of the \df.

The obvious way to constrain the gravitational potential $\Phi$ is to choose
the \df\ that maximises the likelihood of the data for a given potential, and
then to choose the potential that gives the largest likelihood overall. When
\cite{McMillanB13} tried doing this with orbital tori they were thwarted by
the extent of the Poisson noise in their likelihood values.  They found that
the more precise and complete the data in a catalogue are, the less strongly
they constrain the potential because the Poisson noise increases as the data
become more precise. This happens because as each star's error ellipsoid
shrinks, the number of tori that can contribute to the likelihood of that
star decreases. This is a generic problem for any approach to galaxy
modelling that relies on a discrete orbit library. Such approaches include
Schwarzschild modelling \citep[][\S4.7.2]{Schwarzschild,BT08},
made-to-measure modelling \citep[][]{SyerT,DehnenM2M} and straight N-body
modelling. \cite{McMillanB13} showed that the Poisson noise can be
effectively eliminated from the problem if expressions that give
$\vJ(\vx,\vv)$ are available. (By contrast orbital tori yield inverse
expressions $\vx(\vJ,\vtheta)$ and $\vv(\vJ,\vtheta)$.) With such expressions
the scale height and radii of the disc that contributes to $\Phi$ can be
constrained to better than 10\% with a catalogue of just $10\,000$ stars.

\section{Dynamics of the bulge/bar}\label{sec:bulge}

The Galaxy's bulge/bar is one of its three principal components (the other
two being the disc and the dark halo). It is hard to observe, both because it
is heavily obscured by dust and because we see it from within one of its
principal planes. Its dynamics are complex and not fully understood. It has
been modelled with Schwarzschild's technique by \cite{Zhao} and
\cite{Haefner} but the best current models have been obtained by either full
N-body simulation \citep{Martinez11,Martinez13} or the made-to-measure (M2M)
technique \citep{BissantzDG,Long13}. These models have successfully
reproduced both the kinematics of stars observed in regions of low extinction
and the statistics of microlensing events.

On account of our unfavourable location for viewing the Galactic centre,
there was no consensus that ours is a barred galaxy before 1991. Then
\cite{BlitzS} pointed out that near-IR photometry showed the signatures of a
bar seen not far from end-on, signatures that subsequently became very clear
in the photometry from the DIRBE experiment aboard the COBE satellite
\citep{Dwek95,BGS}. This evidence from surface photometry was reinforced by
studies of the luminosity functions of stars seen to the left and right of
the Galactic centre, which showed that stars of a given type seen to the left
of the Galactic centre are systematically brighter than those seen to the
right, presumably because they are nearer \citep{Weinberg92,Stanek}.

\cite{BinneyGSBU} made the case for a central bar on dynamical grounds by
interpreting the longitude-velocity diagrams of CO and \hi\ emission at
longitudes $l\lta12^\circ$ in terms of the two major families of closed
orbits in the equatorial plane of a rotating bar. These are the $x_1$ family,
members of which are elongated parallel to the bar and become more elongated
as one moves inwards, and the $x_2$ family, whose members exist at smaller
radii and are elongated perpendicular to the bar
\cite[][\S3.3.2]{ContopoulosP,BT08}. In the Galaxy the $x_1$ orbits extend
out to radii in excess of $2\kpc$ and gas clouds on these orbits contain
mostly \hi. Clouds on the innermost of the populated $x_1$ orbits are moving
towards us at $\sim54\kms$ as they pass in front of Sgr A*, and form the
``expanding $3\kpc$ arm''. \cite{DameThadeus} were eventually able to
identify emission at positive line-of-sight velocities from clouds on these
orbits as they pass behind Sgr A*, thus conclusively proving that the flow of
gas in the inner few kiloparsecs is governed by the gravitational field of a
rotating bar.  

The $x_2$ orbits are populated by clouds of dense, mostly molecular gas and
are sites of rapid star formation. The Galactic wind \citep{CohenBH} is
probably driven by this star formation. Gas on the outermost $x_2$ orbits at
$R\simeq0.2\kpc$ impacts gas on the innermost $x_1$ orbits, with the result
that gas is constantly transferring from the innermost $x_1$ orbit to the
outermost $x_2$ orbit.  There the gas tends to accumulate, and in external
galaxies a circum-nuclear ring of star formation is often seen \citep{ButaC},
which is presumably the consequence of such gas accumulation.

Traditionally observations of bulge stars have been restricted to windows of
low absorption that lie $\sim4^\circ$ from the plane. In these windows bulge
stars lie $\sim0.5\kpc$ from the plane, so they do not include more recently
formed stars, and, despite the clear evidence of rapid star formation, the
bulge is often considered old.

N-body simulations show that systems like the bulge/bar are the long-term
products of the dynamical evolution of initially cool discs: a planar bar
forms first, and when it has become strong enough and its particles are on
highly eccentric orbits, the bar quite suddenly buckles out of the plane, the
elongation decreases somewhat and a three-dimensional bar/bulge emerges
\citep{CombesS,Raha,MV06}. When viewed from near its minor axis in the plane,
such a bar/bulge has the characteristic peanut shape that is often seen in
external galaxies, and is evident in the 2MASS image of our Galaxy
\citep{Carpenter}. These systems have high pattern speeds in the sense that
their corotation radii lie at about $1.2$ times their major semi-axis lengths
\citep{Athanassoula03}. The pattern speed of the Galactic bar has been
estimated to lie between $40\kms\kpc^{-1}$ \citep{Long13} and
$53\kms\kpc^{-1}$ \citep{DehnenOp}.

The bar/bulge must interact strongly with both the disc and the dark halo.
Its coupling to the stellar disc was explored by \cite{Martinez11}, who
showed that it creates over-densities in the surrounding disc at
$R\simeq4\kpc$ and azimuthal locations that shift from leading the bar to
trailing it
and back again. This phenomenon reflects the fact that the
dominant waves in a disc lie inside their corotation radius, so it is natural
for the pattern of waves in the disc around the bar to rotate more slowly
than the bar \citep{SparkeS}. In fact \cite{MassetTagger} present convincing
evidence that a bar excites a spiral wave in the surrounding disc that is
resonantly coupled to the bar in such a way that it rotates more slowly than
the bar.  Consequently, the density peaks in the disc are constantly
overtaken by the ends of the bar, and we see alternation between leading and
trailing morphology.  This dynamical picture provides an elegant explanation
for observation by \cite{Cabrera08} that clump stars in the disc delineate a
``long thin bar'' that leads the long axis of the ``short-thick bar'' by
$\sim15^\circ$.

\cite{DehnenOp} argued that the Hercules stream, which is a prominent feature
in the velocity-space distribution of nearby stars, comprises stars that
resonate with the bar's pattern speed, and from this conjecture inferred that
the pattern speed is $\Omega_{\rm p}=(53\pm3)\kms\kpc^{-1}$.

The bar must be constantly surrendering angular momentum to the dark halo
\citep{WeinbergT,Sellwood06} and to the stellar disc \citep{MassetTagger}. It
must be acquiring angular momentum from gas which passes through the
corotation resonance and moves on orbits of the $x_1$ family before
transferring to an $x_2$ orbit \citep{Combes10}. In simulations that lack gas,
angular-momentum loss to the dark halo and disc causes the bar to strengthen
and slow down \citep{Athanassoula_flow}. If this process had continued for a
Hubble time, the bar's corotation radius would lie much further out than it
does.  Hence acquisition of angular momentum from gas must be an important
process.

In the vicinity of the corotation resonance, phase space has significant
chaotic zones. That is, many orbits in this region are not quasiperiodic and
do not admit three isolating integrals. For the modeller this fact is a
nuisance because we cannot use Jeans' theorem to construct generic
equilibrium models. But the implications go much deeper than that because
in a chaotic phase space orbits diffuse. In \S\ref{sec:ptheory} we will
present a formalism that allows one to follow the evolution of a
stellar population consequent on the diffusion of its stars through phase
space. However, this apparatus is not immediately applicable to dynamics near
the ends of the bar, because it requires that orbits are essentially
quasiperiodic. How to quantify diffusion through an intrinsically chaotic
phase space is an open question.

\section{The stellar and dark halos}\label{sec:halo-bulge}

We now discuss two components of the Galaxy that have rather similar dynamics
but dramatically different properties in other respects: the dark and stellar
halos.

\subsection{The dark halo}\label{sec:DMhalo}

The dark halo is generally assumed to have a structure similar to that of an
NFW component \citep{NFW}. Such halos form in cosmological simulations of the
clustering of dark matter (\dm) for reasons that are not properly understood.
The most intriguing aspect of the formation of NFW components is that they
arise independent of the power spectrum of the fluctuations in the original
density of \dm\ -- if one suppresses the small-scale power in the fluctuation
spectrum, low-mass halos are eliminated, but the massive halos that remain
are still have central near power-law cusps \citep{Moore99}. These
cusps are a memory of the formally infinite initial phase-space density
of cold dark-matter: as $r\to0$, the density diverges like
$r^{-1}$, so the mass $M(r)$ interior to radius $r$ scales as $r^2$ and the
velocity dispersion scales as $\sigma\propto \sqrt{M/r}\propto r^{1/2}$ and
thus the phase-space density diverges as $\rho/\sigma^3\propto r^{-5/2}$.
Given that the velocity dispersion vanishes with $r$, these cusps are fragile
in the sense that particles within them can be liberated, and thus the cusp
destroyed, by quite a weak scattering event. On the other hand their high
spatial densities ensure that they are not easily tidally disrupted.

The statistics of microlensing events towards the Galactic centre tell us
that inside $r\simeq3\kpc$ the Galaxy is baryon dominated \citep{BissantzDG}.
In fact we may be sure that baryons have been dominant wherever stars have
formed in abundance since gas will not be unstable to gravitational collapse
if the local gravitational field is not dominated by its self gravity.
Dwarf spheroidal galaxies now appear to be dark-matter dominated, but this
must be because at some point dying stars blew away most of the gas from
which those stars formed.

What impact will a period of baryon domination have on  the \dm? Slow
accumulation of baryons, through gas accretion for example, will
adiabatically compress the \dm. In this compression the \df\ $f(\vJ)$
of the \dm\ will be invariant but the distribution of \dm\ in
$(\vx,\vv)$ space will change. \cite{Blumenthal} estimated how \dm\ would
respond to adiabatic compression by considering the trivially computed case
in which all particles are on circular orbits. \cite{SellwoodMcG} treated the
general case and showed that non-zero radial velocity dispersion diminishes
the amplitude of compression.

It is far from clear that adiabatic compression is a useful approximation
because star-forming gas is bound to fragment into massive, dense clumps,
which will efficiently scatter \dm\ particles onto more energetic orbits, thus
decreasing the density of \dm. The physics of this process is that of
dynamical friction -- a gas cloud of mass $m$ that is initially at radius $r$
will surrender its energy to background particles in $\sim M(r)/m$ dynamical
times at $r$, where $M(r)$ is the mass of all material inside $r$. Note that
in an NFW halo the dynamical time $r/\sigma\propto r^{1/2}$ tends to zero
with $r$, so even if many dynamical times were required for modification of
the \dm\ distribution, this would be no problem.  Thus once the gas has come to
dominate the local mass density, only a few dynamical times are required for
the \dm\ profile to be profoundly modified by frictional heating if the gas
fragments into a handful of massive clouds.

The density and velocity distribution of \dm\ near the Sun is important for
experiments that seek to detect \dm\ with cryogenic underground detectors.
Initially cold \dm\ should be restricted to a nearly three-dimensional subset
of six-dimensional phase space: that is, at each spatial point, all \dm\
particles should have essentially the same velocity.  As this ``sheet'' of \dm\
moves through phase space under the influence of gravity, it is stretched and
folded, so that long after virialisation, at any given spatial point many
different folds of the sheet can be found, each with its own characteristic
velocity. In fact the velocity distribution of the \dm\ particles at a point
must be made up of a finite number of contributions at particular velocities
by distinct folds of the sheet. At some spatial points a fold of the sheet is
tangent to a velocity direction $v_i$ so it has a significant velocity
spread. By Poincar\'e's invariant theorem \citep[][\S D.4.2]{BT08}, the wide
spread in velocity of this fold must be compensated by a narrow spread in the
canonically conjugate spatial coordinate $x_i$, and the spatial density of
this fold may be orders of magnitude larger than that of a generic fold.  How
irregular will the velocity distribution of the \dm\ be on Earth, and can folds
of exceptional density significantly enhance the rate of self-annihilation by
\dm?  \cite{VogelsbergerWhite} used adapted cosmological N-body simulations to
address these questions. They concluded that very large numbers
($\sim10^{14}$) of folds contribute to the velocity distribution at the Sun,
with half of the \dm\ being contributed by the $\sim10^6$ most massive folds.
They concluded that folds of anomalous spatial density do not dominate the
total annihilation rate. Thus the naive model in which the \dm\ has a smooth
\df\ $f(\vx,\vv)$ should be able to account for all observable phenomena.
 
\subsection{The stellar halo}\label{sec:stellarhalo}

In \S\ref{sec:haloDF} we discussed the choice of \df s to represent the
stellar halo. The idealisation of a smooth halo, which these \df s embody, is
often useful even though we now know that the stellar halo is far from smooth
-- \cite{Bell} report that at least half of the halo's luminosity is
accounted for by lumps and streams. Thus the stellar halo is markedly less
smooth than the dark halo is thought to be, presumably because by radiating
binding energy baryons early on form more tightly bound clumps at each mass
scale than (dissipationless) \dm\ can.

A classical application of the
concept of a smooth stellar halo is to the estimation of the escape speed
$v_{\rm esc}$ from the solar neighbourhood \citep{SmithEscape}.  Essentially
one counts fast stars and following \cite{LeonardTremaine} fits the number
counts to a formula based on the assumption that the \df\ is of the form
$f(E)$ and is such that $f\to0$ as $E\to0$. The weakness of this procedure is
that as $E\to0$ the periods of orbits tend to infinity.  Consequently, in the
limit $E\to0$, the time for the system to phase mix tends to infinity, and we
cannot invoke Jeans' theorem.  In fact, in this limit we must expect the \df\
to depend on the angle variables as well as on the integrals. Moreover, there
{\it will\/} be a population of unbound, escaping stars, so the assumption
that $f=0$ for $E>0$ is also unsafe. The determination of $v_{\rm esc}$
involves extrapolation of the data, which is always dangerous. However,  in
so far as simulations of galaxy formation are realistic, it makes sense to
estimate the parameters of the Galaxy's potential, by fitting the observed velocity
distribution of fast stars to simulations \citep{Piffl13}

\subsection{Tidal streams}\label{sec:streams}

The SDSS revealed several long thin over-densities of stars
\citep{Grillmair06,BelokurovFoS}. The most prominent of these is clearly the
tidal debris of the Sagittarius dwarf spheroidal galaxy
\citep{Majewski,Fellhauer} and SDSS photometry of the vicinity of the
globular cluster Palomar 5 allows us to study the process of disruption in
detail \citep{Newberg02,Odenkirchen09}. Consequently, all linear
over-densities are assumed to be generated by the tidal shredding of either a
star cluster or a dwarf galaxy, even though some streams have no known
progenitor.

When a stream's progenitor passes through pericenter, the Galaxy's tidal
field detaches some stars from the point nearest the Galactic centre, and
other stars from the furthest point. The stars detached from the near point
have less angular momentum and shorter-period orbits than the progenitor,
while the stars detached from the far point have more angular momentum and
longer periods than the progenitor. When they are first stripped, all stars
have similar phases, but over time the spread in the phases of stripped stars
increases because the phases of stars detached from the near point advance
more rapidly than do the phases of stars detached from the far point.

Mathematically, let $\delta\vJ$ be the difference between the actions of a
detached star and the progenitor's actions. Then the frequencies $\vOmega$ of
the detached star differ from those of the progenitor by 
 \[
\delta\vOmega=\vD\cdot\delta\vJ,
\]
 where
\[
D_{ij}={\pa^2H\over\pa J_i\pa J_j}={\pa\Omega_j\over\pa J_i}
\]
 is the Hessian matrix. $\vD$ is a symmetric matrix, so it can be expanded 
as a sum over its orthogonal eigenvectors $\ve_i$ and (real) eigenvalues
$\lambda_i$:
 \[\label{eq:ketex}
\vD=\sum_i\lambda_i\ve_i\otimes\ve_i.
\]
 \cite{TremaineStream} pointed out that for typical Galactic potentials $\vD$
is a very anisotropic matrix in that one of its eigenvalues, $\lambda_0$, is
much bigger than the others. This eigenvalue will dominate the sum
(\ref{eq:ketex}), so to a good approximation we have
 \[
\delta\vOmega\simeq\lambda_0\ve_0(\ve_0\cdot\delta\vJ).
\]
 A long time $t$ after the star was detached from the progenitor, its
displacement in angle space from the location of  the progenitor is
 \[\label{eq:thetaOmega}
\delta\vtheta\simeq\delta\vOmega
t\simeq\lambda_0t(\ve_0\cdot\delta\vJ)\,\ve_0,
\]
 where we have neglected the small initial displacement. Thus in angle space
the stream is drawn out into a straight line parallel to $\ve_0$ -- the speed
at which a star moves along this line is proportional to
$\delta\vJ\cdot\ve_0$. This straight line in angle space maps into a curve in
real space. Thus streams form because their stars are {\it not\/} all on the
same orbit but on orbits that vary systematically along the stream, contrary
to what was for many years assumed
\citep{JohnstonStream,BinneyStream,EyreBI,EyreBII}. In angle space the
progenitor's orbit is the straight line parallel to $\vOmega$ evaluated at
the progenitor's actions. \cite{EyreBIII} showed that in the (spherical)
isochrone potential the angle between $\ve_0$ and $\vOmega$ is typically
$1-2$ degrees, and \cite{JasonI} show that in more realistic potentials the
misalignment can exceed $10^\circ$. Moreover, even an angle difference
$\sim2^\circ$ can lead to large errors in the derived gravitational potential
when
a potential is sought that makes an orbit run along a stream.

\cite{JasonII} propose an alternative way to diagnose the potential with
stream data and test its ability to infer correctly the flattening $q$ and
circular speed $V_c$ of the flattened logarithmic potential in which a stream
has been constructed by tidal disruption of a self-gravitating N-body model.
For each trial potential one plots the stars of the stream in both angle
space and frequency space. On account of the first equality of equation
(\ref{eq:thetaOmega}) the lines of regression through these two distributions
should be parallel. One seeks potentials in which this condition is satisfied
to within the errors. With data comparable to those that will be furnished by
Gaia, the errors on the derived values of $V_c$ are $\sim4\kms$ and those on
$q$ $\sim6\%$.

\section{Evolution of the potential}\label{sec:ptheory}

To this point we have been modelling the Galactic potential as a smooth,
time-independent object. In reality the potential contains a significant
fluctuating component caused by moving gas clouds, stellar systems and lumps
of dark matter, and spiral structure. The potential also evolves secularly in
response to the infall of gas, stars and dark matter, becoming ever deeper
and probably flatter. Changes in the potential give rise to changes in stars'
orbits. We now discuss such changes. 

A fundamental result is obtained by multiplying the equation of motion
$\dot\vv=-\nabla\Phi$ by $\vv$ and rearranging the result to
 \[
{\d E\over\d t}={\pa\Phi\over\pa t}.
\]
 Thus stars change their energies if and only if the potential is
time-dependent.  Fluctuations in the potential redistribute energy between
stars. The overall effect of this redistribution will be to increase the
system's entropy by making it hotter and more centrally concentrated
\citep[][\S4.10.1]{BT08}.

Changes in the potential that occur on a timescale that is significantly
longer than the longest orbital timescale are easy to deal with because such
changes leave the actions $\vJ$ invariant \citep[][\S3.6]{BT08}. This fact greatly
facilitates the process of determining the response of a stellar system to
adiabatic distortion of its potential because all we have to do is to move
each star from  its orbit in the original potential to the orbit with the
same actions in the distorted potential. In particular, the structure of the
distorted system depends only on the initial and final configurations, and
not on which configurations it passed through in between.

The issue of when a change is slow enough to leave $\vJ$ invariant is a
tricky one because there are infinitely many timescales
$t_\vn\equiv2\pi/|\vn\cdot\vOmega|$ associated with a given orbit, where
$\vn$ is any vector with integer components. For sufficiently large $|\vn|$,
$t_\vn$ can be arbitrarily long, and even a tiny perturbation can in
principle lead to violation of adiabatic invariance through the mechanism of
resonant trapping that we described qualitatively in \S\ref{sec:int_res}. We
now analyse this phenomenon mathematically.

\subsection{Resonant trapping}\label{sec:trapping}

Let $h(\vtheta,\vJ)$ be the difference between the true $H$ and a nearby integrable
Hamiltonian $H_0(\vJ)$, which might be one constructed by the torus machine
(\S\ref{sec:intH}).  Hamilton's equations for motion in $H$ read
 \[
\dot\vtheta={\pa H\over\pa\vJ}=\vOmega_0+{\pa h\over\pa\vJ},
\qquad
\dot\vJ=-{\pa H\over\pa\vtheta}=-{\pa h\over\pa\vtheta},
\]
 where $\vOmega_0=\pa H_0/\pa\vJ$.  The perturbing Hamiltonian, like any
function on phase space, is a periodic function of the angles, so we can
Fourier expand it: 
 \[
h(\vtheta,\vJ)=\sum_\vn h_\vn(\vJ)\cos(\vn\cdot\vtheta+\psi_\vn).
\]
 With this expansion, the equation of motion of $\vJ$ is
 \[
\dot\vJ=\sum_\vn \vn h_\vn(\vJ)\sin(\vn\cdot\vtheta+\psi_\vn) 
=\sum_\vn \vn h_\vn(\vJ)\sin(\vn\cdot\vOmega\,t+\psi'_\vn),
\]
 where $\psi_\vn'=\psi_\vn+\vn\cdot\vtheta(0)$. So long as
$\vn\cdot\vOmega\ne0$, the time-averaged value of $\dot\vJ$ vanishes and we
expect $\vJ$ simply to oscillate slightly around its unperturbed value.  But
if a \deffn{resonance condition} $\vN\cdot\vOmega=0$ is nearly satisfied, the
argument of one or more of the sines will change very slowly, and the
cumulative change in $\vJ$ can be significant even for very small $h_\vN$.
When we multiply the resonance condition by $t$, we obtain
 \[
\vN\cdot\vtheta(t)=\const.
\]
 This equation inspires us to make a canonical transformation to new angles
and actions $(\vtheta',\vJ')$ such that $\vN\cdot\vtheta$ becomes one of the
new angles, say $\theta'_1$.  It does not much matter what we adopt for
$\theta'_2$ and $\theta'_3$; $\theta'_2=\theta_2$ and $\theta'_3=\theta_3$
works fine. Then the generating function of the required transformation is
 \[\label{eq:resS}
S(\vtheta,\vJ')=J'_1\vN\cdot\vtheta+J'_2\theta_2+J'_3\theta_3,
\]
 and the new angles are
 \[\label{eq:restheta}
\theta'_1={\pa S\over\pa J'_1}=\vN\cdot\vtheta,\quad
\theta'_2={\pa S\over\pa J'_2}=\theta_2,\quad
\theta'_3=\theta_3.
\]
 Since $\theta'_1$ does not evolve in time, the star explores only a
two-dimensional set of its three-dimensional torus. While a star on a
resonant torus does not come arbitrarily close to every point on its torus,
the bigger the numbers $N_j$ are, the more effectively it samples its torus,
and the less likely it is that the resonance condition will be dynamically
important.

Mathematically, we use the new angle variables $\vtheta'$ defined by equation
(\ref{eq:restheta}) and their conjugate actions $\vJ'$, which follow from
\[
J_1={\pa S\over\pa\theta_1}=J'_1N_1,\quad
J_2={\pa S\over\pa\theta_2}=J'_1N_2+J'_2,\quad
J_3=J'_1N_3+J'_3.
\]
 Thus
\[
J_1'=J_1/N_1\quad;\quad J_2'=J_2-J_1N_2/N_1\quad;\quad
J_3'=J_3-J_1N_3/N_1.
\]
 Next we express $h$ as a Fourier series in the new angle variables and
discard all terms that involve $\theta'_2$ or $\theta'_3$ on the grounds that
they oscillate too rapidly to have a significant impact on the dynamics.
Since our approximated Hamiltonian depends on neither $\theta'_2$ nor
$\theta'_3$, to the level of our approximations the conjugate actions $J'_2$
and $J'_3$ will be constants of motion. The only non-trivial equations of
motion are now
 \[\begin{split}\label{eq:slowmotion}
\dot \theta'_1&=\Omega'_{01}(J'_1)+\sum_n{\pa h'_{(n,0,0)}\over\pa
J'_1}\cos(n\theta'_1+\psi'_{(n,0,0)})\\
\dot J'_1&=\sum_n nh'_{(n,0,0)}(J'_1)\sin(n\theta'_1+\psi'_{(n,0,0)}),
\end{split}\]
 where 
\[
\Omega'_{01}={\pa H_0(\vJ')\over\pa J'_1}=\sum_i\Omega_{0i}{\pa J_i\over\pa J'_1}
=\Omega_{01}N_1+\Omega_{02}N_2+\Omega_{03}N_3=\vN\cdot\vOmega_0
\]
 and we have suppressed the dependence of $\Omega_{01}$ and $h'_{(n,0,0)}$ on
$J'_2$ and $J'_3$ because the latter are mere constants.
We have reduced the particle's motion in six-dimensional phase space to
motion in the $(\theta'_1,J'_1)$ plane.

We differentiate the first of equations
(\ref{eq:slowmotion})  with respect to time
 \[
\ddot\theta'_1={\pa\Omega'_{01}\over\pa J'_1}\dot J'_1+\sum_n
\biggl(
{\pa^2 h'_{(n,0,0)}\over\pa {J'_1}^2}\dot J'_1\cos(n\theta'_1+\psi'_{(n,0,0)}
)-
n\dot\theta'_1{\pa h'_{(n,0,0)}\over\pa J'_1}\sin(n\theta'_1+\psi'_{(n,0,0)}
)\biggr).
\]
 We can neglect the sum in this equation because each of its terms is the
product of a derivative of $h'_{(n,0,0)}$, which is small, and either $\dot J'_1$,
which is of the same order, or $\dot\theta'_1$, which is also small because
$\Omega'_{01}$ vanishes at the resonance. Therefore we can dramatically
simplify the $\theta'_1$ equation of motion to
\[\label{eq:genpend}
\ddot\theta'_1={\pa\Omega'_{01}\over\pa J'_1}\dot J'_1
={\pa\Omega'_{01}\over\pa J'_1}\sum_n nh'_{(n,0,0)}(J'_1)\sin(n\theta'_1+\psi'_{(n,0,0)}
).
\]
 If we approximate $\pa\Omega'_{01}/\pa J'_1$ and $h'_{(n,0,0)}$ by their values on
resonance, and retain only the term for $n=1$ in the sum, we are left with the
equation of motion of a pendulum
 \[\label{eq:pendule}
\ddot\theta=-\omega^2\sin\theta,
\]
 where
 \[
\theta\equiv\theta'_1+\psi'_{(1,0,0)} \quad\hbox{and}\quad
\omega^2\equiv-{\pa\Omega'_{01}\over\pa J'_1}h'_{(1,0,0)}.
\]

\begin{figure}
\centerline{\includegraphics[width=.8\hsize]{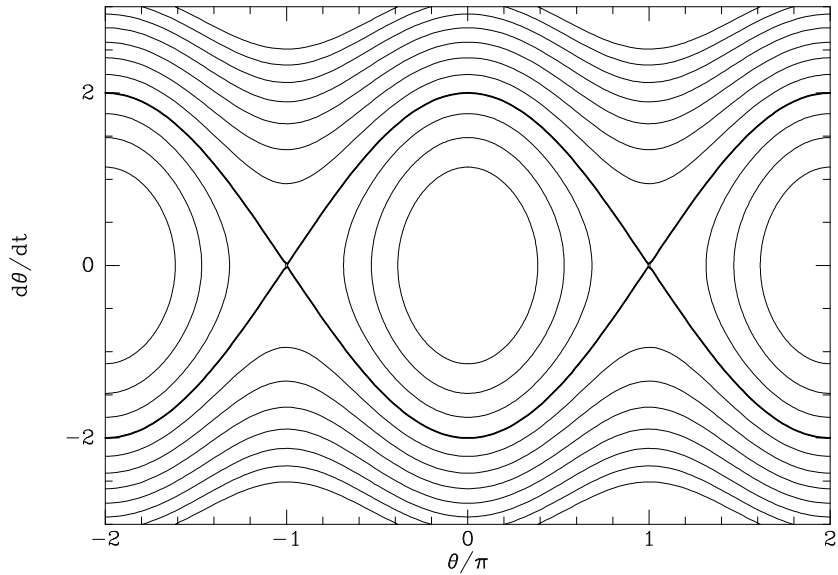}}
\caption{The phase plane of a pendulum.  Curves of constant energy $E$
(eq.~\ref{eq:Epend}) are plotted. The particle moves on these, from left to
right in the upper half of the figure and from right to left in the lower
half.}\label{fig:pendule}
\end{figure}

A pendulum can move in two ways: at low energy its motion is
oscillatory at an angular frequency that falls from $\omega$ at the lowest
energies to zero at the critical energy above which the pendulum circulates.
Consequently, equation (\ref{eq:genpend}) predicts that close to the resonance
(``low energy'') $\theta'_1$ will oscillate. This is the regime of
{resonant trapping} in which the particle \deffn{librates} around the
underlying resonant orbit. At some critical distance from the resonance
(``high energy'') $\theta'_1$ will start to circulate. 
We obtain a useful pictorial representation of this behaviour by
deriving the energy invariant of equation (\ref{eq:pendule}). We multiply
both sides by $\dot\theta$ to obtain an equation which states that
 \[\label{eq:Epend}
E\equiv\fracj12\dot\theta^2-\omega^2\cos\theta
\]
 is constant. Consequently, the particle moves in the $(\theta,\dot\theta)$
plane along curves of constant $E$ like those shown in \figref{fig:pendule}.
The round contours near the centre of the figure show the motion of a
particle that has been trapped by the resonance, while the contours at the
top and bottom of the figure, which run all the way from left to right, show
the motion of a particle that continues to circulate.

Near the threshold energy, the phase point moves close to one of the heavy
curves in the figure, and the rate at which $\theta'_1$ advances in time is
highly non-uniform, just as a pendulum that has only just enough energy to
get over top dead centre slows markedly as it does so. As we move further and
further from the resonance and either up beyond the figure's top boundary
or down below its bottom boundary, the rate of advance of $\theta'_1$ becomes more
and more uniform, and we gradually recover the unperturbed motion, in which
the rate of advance of $\theta'_1$ is strictly uniform.

\subsubsection{Levitation}\label{sec:levitaton}

\begin{figure}
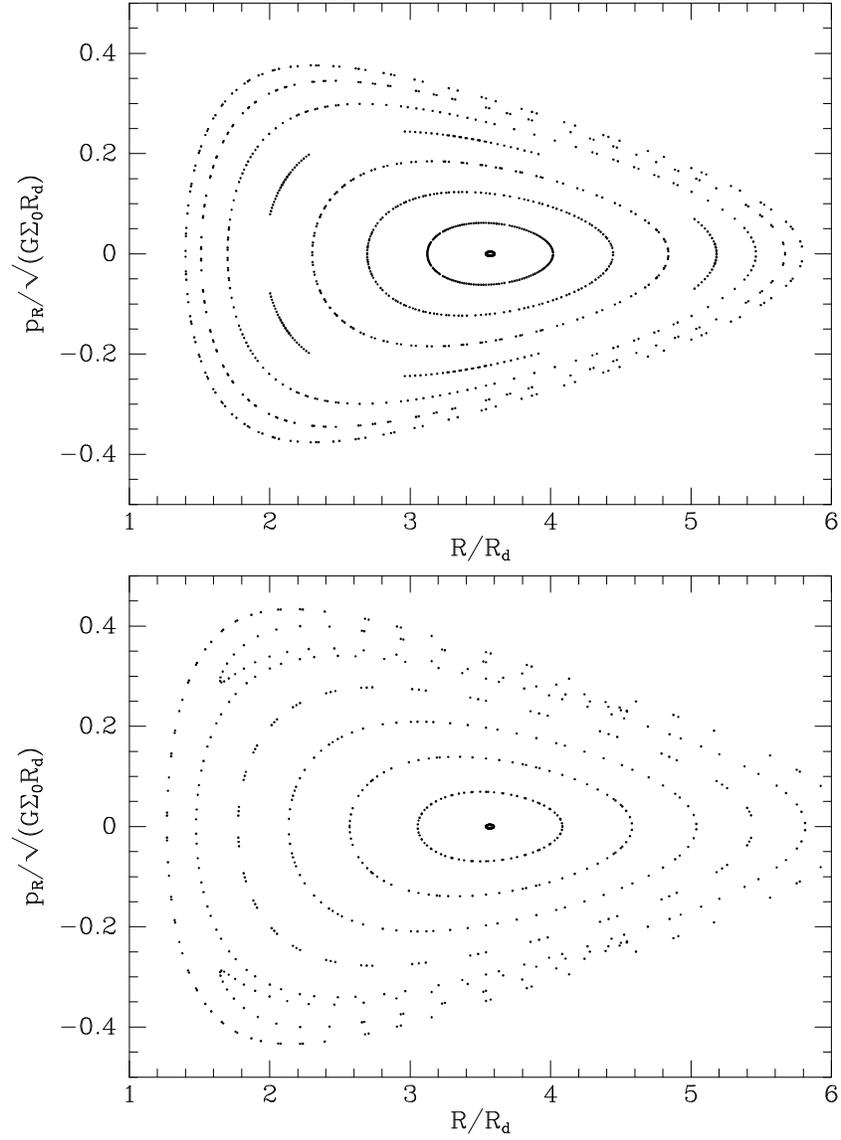

\centerline{\includegraphics[width=.8\hsize]{fig5a.ps}}\vskip5pt
\centerline{\includegraphics[width=.8\hsize]{fig5b.ps}}
\caption{Surfaces of section for motion in flattened isochrone
potentials: the upper panel is for the case of a mass distribution that has
axis ratio $q=0.7$, while the lower panel is for $q=0.4$. In the lower panel
we see resonant islands generated by the $1:1$ resonance between the radial and
vertical oscillations. No such island is evident in the upper panel.}
\label{fig:resonance}
\end{figure}

Resonant trapping shows up clearly in a surface of section such as those
shown in \figref{fig:resonance} for motion in two axisymmetric potentials
$\Phi(R,z)$.  Each orbit gives rise to a series of points that lie on the
curve in which the orbit's torus intersects the $(R,p_R)$ plane. All orbits
contributing to a given panel of \figref{fig:resonance} have the same energy
and value of $L_z$. Most of the curves move around a central point. The point
itself is made by the \deffn{shell orbit} $J_r=0$; in real space this orbit
is a thin cylindrical shell that has a larger diameter at $z=0$ than at its
top or bottom edges. Each curve around this point in \figref{fig:resonance}
is generated by a three-dimensional orbit that forms an annulus of finite
thickness. In \figref{fig:resonance}, the longer an orbit's curve is, the
thicker the real-space annulus and the smaller its vertical extent. The
outermost curve in \figref{fig:resonance} is generated by the orbit $J_z=0$,
which is confined to the plane $z=0$.

The lower panel in \figref{fig:resonance} is for motion in the potential of a
flatter galaxy than the upper panel, and in this panel not all curves loop
around the central point. Two \deffn{resonant islands} have appeared, formed
by orbits that have been trapped by the resonance $\Omega_r=\Omega_z$.

Before the disc formed, when the Galaxy's potential was nearly spherical,
every orbit had a value of $\Omega_r$ that was bigger than either
$\Omega_\phi$ or $\Omega_z$. Stars whose orbits are confined within a couple
of kiloparsecs of the equatorial plane now have $\Omega_r<\Omega_z$. So at
some point in the flattening process these stars satisfied the resonance
condition $\Omega_r=\Omega_z$.  In a flattened potential $\Omega_r/\Omega_z$ is
smallest for orbits that are confined to the equatorial plane. Therefore the
resonance condition was first satisfied by these orbits.

\figref{fig:resonance} shows surfaces of section for motion in a potential
before and after the resonance condition $\Omega_r=\Omega_z$ is first
satisfied: the islands visible in the lower panel are made up of orbits trapped
by this resonance.  Note that the areas of the curves that loop around the
central point are
 \[
\int \d R\,\d p_R=2\pi J_r,
\]
 so these areas do not change as the potential flattens.

The resonance condition is first satisfied by the orbit that is confined to
the equatorial plane; in both panels of \figref{fig:resonance} the curve of
this orbit lies on the outside. Hence the resonant islands first appeared
just inside this curve. As the potential flattened more, $\Omega_r/\Omega_z$
dropped significantly below unity for the planar orbit, so the resonance
condition was satisfied by orbits with non-zero $J_z$ and the islands moved
inwards. Orbits whose curves lay in the path of a moving island did one of
two things \citep[][\S3.7.2]{BT08}: (a) they were trapped into the island, or
(b) they abruptly increased their radial actions so that their curves went
round the far side of the island.  Which of these two outcomes happened in an
individual case depended on the precise orbital phase of the star when the
potential achieved a particular flattening, but it is most useful to average
over phases and to consider the outcomes to occur with probabilities $P_{\rm
a}$ or $P_{\rm b}$. The magnitudes of $P_{\rm a}$ and $P_{\rm b}$ depend of
the relative speed with which the island increased its area and moved: if it
simply grew, $P_{\rm a}=1$, and if it moved without growing $P_{\rm b}=1$.
These results follow from Poincar\'e's invariant theorem, which implies that
the Hamiltonian flow shuffles particles around surfaces of section without
changing the density of particles \citep[for some subtleties in this
see][]{BinneyGH}.

Let's imagine that after a period of stationary growth, the island moved
inwards without growing, and then became stationary while it shrank. In this
case it would have swept up stars with large $J_r$ and small $J_z$ and
released these stars into orbits with smaller $J_r$ and larger $J_z$. In
other words, it will have turned radial motion into vertical motion.
\cite{Touma} have called this process ``levitation''.  Conversely, the moving
island will have reduced the vertical motions and increased the radial
motions of any stars it found in its path through action space. Thus
resonances stir the contents of phase space. Levitation is a lovely idea but
it's not clear that it is of practical importance. In the next section we
shall see that in a disc analogous scattering by resonances is very
important.

\begin{figure}
\centerline{
\includegraphics[width=.75\hsize]{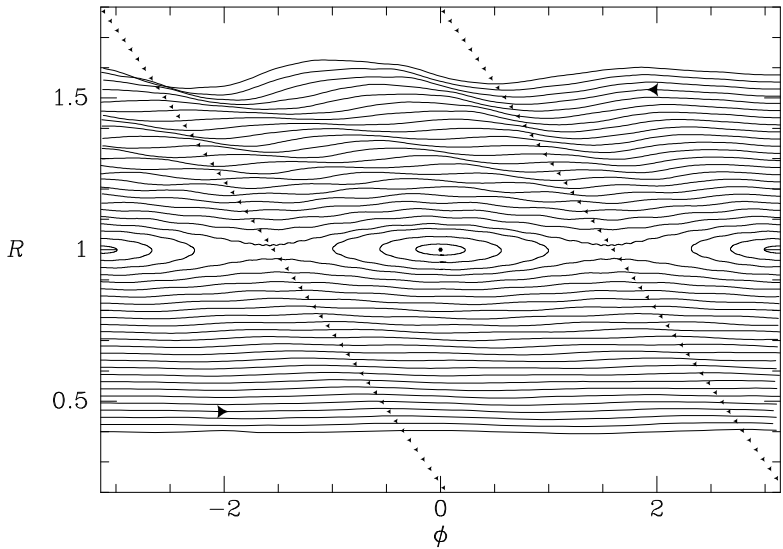}}
\caption{The response of initially circular orbits to a
spiral potential (after Sellwood \& Binney 2002).}\label{fig:SellB}
\end{figure}

\subsubsection{Stellar migration}\label{sec:migration}

When the pattern speed $\Omega_{\rm p}$ of a non-axisymmetric perturbation
lies near the azimuthal frequency $\Omega_\phi$ of an orbit, the time
$\pi/|\Omega_\phi-\Omega_{\rm p}|$ during which the perturbation torques the
orbit in the same sense is long. Consequently, in these circumstances even a
weak perturbation can induce non-negligible changes in the orbit's value of
$L_z$. Since $L_z$ determines an orbit's guiding-centre radius, changes in
$L_z$ are associated with radial migration.

\figref{fig:SellB} shows the effect of a weak spiral perturbation on
originally circular orbits in a typical Galactic potential. Orbits of small
radius (bottom of the figure) have $\Omega_\phi>\Omega_{\rm p}$ and overtake
the structure, while orbits with large radii have $\Omega_\phi<\Omega_{\rm
p}$ and are overtaken by the structure. Orbits for which
$\Omega_\phi\simeq\Omega_{\rm p}$ so they almost corotate with the structure,
become resonantly trapped by it: on these \deffn{horse-shoe orbits} a star
oscillates around the corotation radius $R_{\rm CR}$ (where
$\Omega_\phi=\Omega_{\rm p}$) -- the ellipses at the centre of \figref{fig:SellB}
show these orbits. During the left half of a horse-shoe orbit, the perturbing
force on the star is dominated by the potential trough that is marked by the
left sloping dashed line, so the star is pulled to the left and its angular
momentum diminishes.  The star moves inwards in response.  Conversely, on the
right half of the orbit, the net perturbative force is to the right, $L_z$
increase, and the star moves outwards. The orbit librates around the trapping
closed orbit that is marked by the dot.\footnote{The spatial location of the
trapping orbit is one of the Lagrange points $L_4/L_5$.}

Now suppose the perturbation grows in amplitude from zero, peaks and fades
away again. As the strength of the perturbation grows, the range of orbits
around corotation that become trapped increases, and the scale of the largest
ellipse in \figref{fig:SellB} grows. When the perturbation fades, stars are
progressively released from entrapment, those on the largest ellipses first.
The stars on a given ellipse will be a mix of stars that joined the ellipse
on its upper half ($R>R_{\rm CR}$) and ones that joined the ellipse on its
lower half. During the lifetime of the perturbation stars circulate around
the ellipse, so it's unclear whether a star will be released from the top or
the bottom of the ellipse.  Roughly half the stars that became trapped when
$R<R_{\rm CR}$ are released at $R>R_{\rm CR}$: the guiding-centre radii of
these stars have been permanently increased. Conversely, roughly half the
stars that were at $R>R_{\rm CR}$ when they became trapped are released at
$R<R_{\rm CR}$.  Thus a transient non-axisymmetric perturbation causes stars
in a zone around the corotation radius to migrate radially either inwards or
outwards.

\begin{figure}
\centerline{\includegraphics[width=.75\hsize]{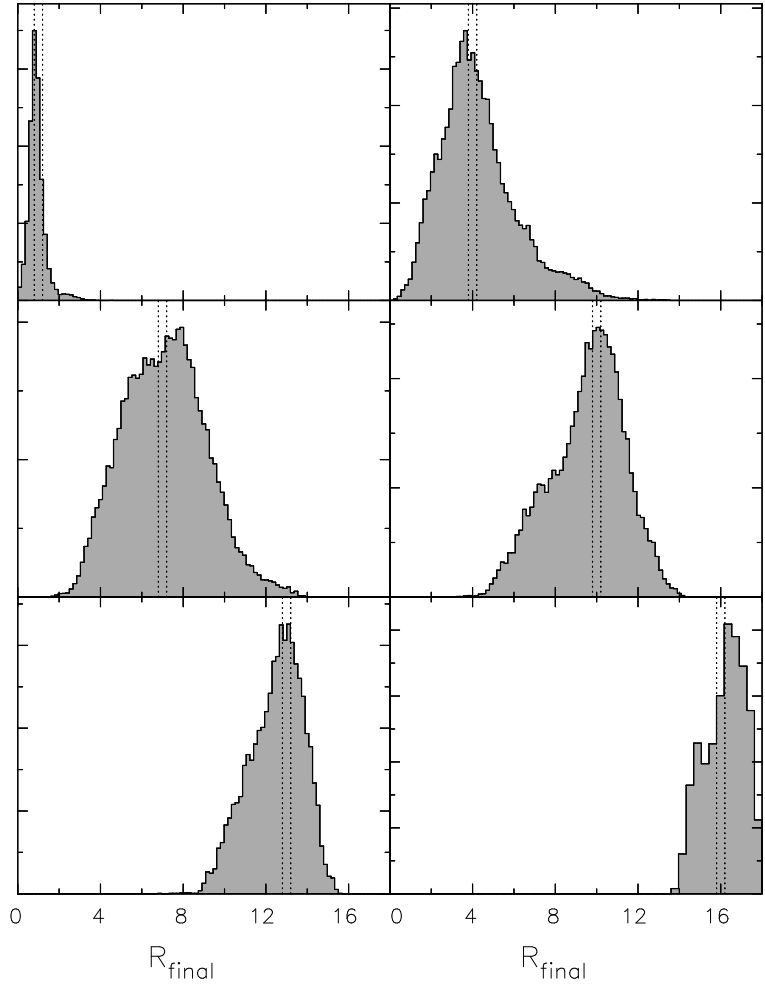}}
 \caption{Each panel shows the distribution of the final guiding-centre radii
of the stars in a disc simulation whose initial guiding-centre radii lay
within the region between the dotted vertical lines. The disc had a flat
rotation curve, $Q=1.5$, and half of the radial force was provided by a fixed
halo. The duration of the simulation was $\sim4\Gyr$. (From Sellwood \&
Binney 2002).}\label{fig:beforeafter}
\end{figure}

Spiral structure is a natural source of transient non-axisymmetric
perturbations. \cite{SellwoodB} showed that spiral features in N-body
simulations cause a significant level of migration. \figref{fig:beforeafter}
shows histograms of the final radii of stars that all started from the narrow
radial bands marked in an N-body disc that was evolved for $\sim5\Gyr$ in the
absence of any seeded spiral structure. Irregular and quite weak spiral
structure emerged as the disc was evolved.  Stars that started at the current
radius of the Sun finished at radii that are often $1.2-2\kpc$ from $R_0$.
From the strength of spiral structure seen in NIR photometry by \cite{RixZ}, Sellwood \&
Binney estimated that over the Hubble time stars will typically migrate
$\sim2\kpc$ from their birth radii.

Although \figref{fig:SellB} is plotted for circular orbits, the mechanism of
radial migration works also for stars on mildly eccentric orbits.
\cite{SolwaySS} have shown that the effectiveness of radial migration
declines only slowly with increasing eccentricity of the initial orbits.

As we will discuss in \S\ref{sec:spirals}, satellites, including completely
dark ones, that are on eccentric orbits are a natural source of transient
non-axisymmetric perturbations that complement spiral structure. At
pericentre a satellite is moving faster than the local circular speed, so the
corotation radius of its perturbation lies inside pericentre, and thus may
lie in the body of the disc even if pericentre lies outside it.

Any non-axisymmetric perturbation that either does not last for ever, or
changes its pattern speed, is likely to induce radial migration. Bars seem to
be long-lived structures, but, as we saw in \S\ref{sec:bulge}, their pattern
speeds evolve in response to changes in the angular momentum of the
population that supports the bar, and acquisition of angular momentum from
gas, and loss to stars and \dm\ particles, are likely to be constantly
adjusting $\Omega_{\rm p}$.  We return to the issue of bar-driven migration
in \S\ref{sec:barmigrate}.

\begin{figure}
\centerline{\includegraphics[width=.6\hsize]{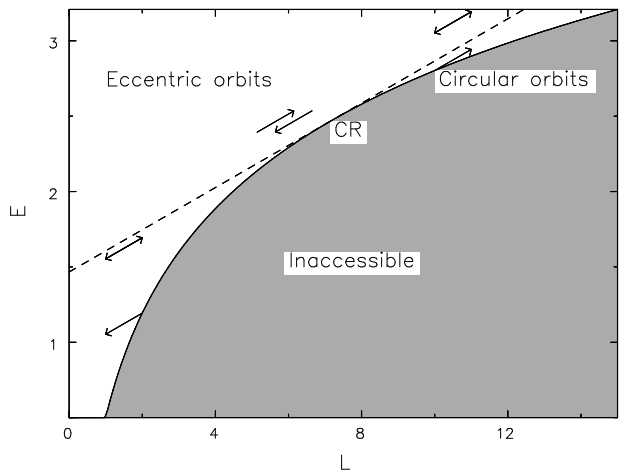}}
 \caption{Energy
versus angular momentum for planar orbits in an axisymmetric potential -- a
``Lindblad diagram''. No orbits lie in the shaded area, which is bounded by
the points of circular orbits. A potential that is stationary in a rotating
frame moves stars along lines with slope $\d E/\d L_z=\Omegap$. (From
Sellwood \& Binney 2002)}
\label{fig:Lindblad}
\end{figure}

\subsubsection{Heating versus migration}\label{sec:heatingmigraton}

Let's assume that a non-axisymmetric perturbation has a nearly constant
pattern speed $\Omegap$. If we work in the frame of reference that rotates at
frequency $\Omegap$, the motion of each star is governed by a
time-independent Hamiltonian, the numerical value of which, the \deffn{Jacobi
constant}, is an isolating integral. In terms of the energy $E$ of motion in
the non-rotating frame, the rotating-frame Hamiltonian is
 \[
H=E-\Omegap L_z.
\]
 Since $H$ is an integral, along an orbit, $\d H=0$ and changes in $E$ and $L_z$ caused by
the perturbation satisfy
 \[\label{eq:Lindblad}
\d E=\Omegap\,\d L_z.
\]

 \figref{fig:Lindblad} is a plot of $E$ versus $L_z$ for a planar
axisymmetric potential. The physically accessible region is bounded below by
the locus of circular orbits, which are the orbits with the largest value of
$L_z$ for each given $E$, so there are no orbits in the shaded region below
this boundary. The slope of the boundary, $(\pa E/\pa L_z)_{J_r=0}$, is the
circular frequency $\Omega(L_z)$. Equation (\ref{eq:Lindblad}) states that a
steadily rotating perturbation moves stars in this figure on lines of slope
$\Omegap$ as indicated by arrows in \figref{fig:Lindblad}. These lines are
tangent to the bounding curve at the {corotation resonance} (\CR), where
$\Omega(L_z)=\Omegap$. Thus at \CR\ the perturbation scatters stars along the
boundary, from one circular orbit to another. Elsewhere, the perturbation
scatters stars away from the boundary, to places where the energy exceeds
that of the circular orbit of the given value of $L_z$, and the additional
energy will be invested in epicyclic motion. Inside the \CR, the angular
momenta of stars that are initially on circular orbits must be reduced, while
outside the \CR\ it must be increased.  Thus perturbation of a cold disc must
move angular momentum outwards.

In summary: away from \CR\ in a cold disc stars have to migrate inwards at
$R<R_{\rm CR}$ and outwards at $R>R_{\rm CR}$. In either case, migration will
inevitably increase the stars' in-plane velocity dispersion, that is, the
perturbation will heat the disc by transferring angular momentum outwards --
galactic discs are giant accretion discs. Near $R_{\rm CR}$ stars can move
either in or out from circular orbit to circular orbit, so migration is
reversible and occurs without heating.

\subsection{Orbit-averaged Fokker-Planck equation}\label{sec:FP}

Our discussion of time-dependent perturbations has so far concentrated on
approximately steadily rotating potentials. Significant effects are caused by
perturbations that do  not simply rotate rigidly, for example the
gravitational fields of massive objects on non-circular orbits. To understand
the impact of such perturbations we now develop a general framework for
handling the impact of fluctuations. The general idea is that, by the strong
Jeans theorem the galaxy's distribution function is at all times a function
$f(\vJ,t)$ of the actions. Fluctuations (and resonances) cause this function
to evolve by causing innumerable small changes in the actions of
individual stars. Let $P(\vJ,\vDelta)\d^3\vDelta\,\delta t$ be the
probability that in time $\delta t$ a star with actions $\vJ$ is scattered to
the action-space volume $\d^3\vDelta$ centred on $\vJ+\vDelta$. The number of
stars in the action-space volume $\d^3\vJ$ is $(2\pi)^3f(\vJ,t)\d^3\vJ$, so
the number of stars leaving this volume in $\delta t$ is
\[
(2\pi)^3f(\vJ,t)\d^3\vJ\delta t\int\d^3\vDelta\, P(\vJ,\vDelta).
\]
 Similarly, the number of stars that are scattered {\it into\/} this volume
 is
\[
(2\pi)^3\d^3\vJ\delta t\int\d^3\vDelta\, f(\vJ-\vDelta,t)P(\vJ-\vDelta,\vDelta).
\]
 Hence the rate of change of the distribution function is
\[\label{eq:master}
{\pa f\over\pa
t}=\int\d^3\vDelta\,\big[f(\vJ-\vDelta,t)P(\vJ-\vDelta,\vDelta)-f(\vJ,t)P(\vJ,\vDelta)
\big].
\]
 Since scattering events change actions only slightly,
$P(\vJ,\Delta)$ is appreciable only for $|\vDelta|\ll|\vJ|$. So we can
truncate after just a few terms the Taylor series expansion in $\vJ$ of the product
$f(\vJ,t)P(\vJ,\vDelta)$:
 \begin{equation}
\begin{split}
f(\vJ-\vDelta,t)P(\vJ-\vDelta,\vDelta)&=f(\vJ,t)P(\vJ,\vDelta)\\
&\quad-\Delta_i{\pa(fP)\over\pa J_i}+\fracj12\Delta_i\Delta_j{\pa^2(fP)\over\pa J_i\pa
J_j}+\cdots
\end{split}\end{equation}
 Substituting the first three terms on the right side of  this expression
into equation (\ref{eq:master}) and cancelling terms, we obtain
 \[\label{eq:FP}
{\pa f\over\pa t}=-{\pa F_i\over\pa J_i},
\quad\hbox{where}\quad
F_i\equiv f\overline{\Delta_i}-\fracj12{\pa(f\overline{\Delta^2_{ij}})\over\pa
J_j},
\]
 \[\label{eq:diffcoeffs}
\overline{\Delta_i}(\vJ)\equiv\int\d^3\vDelta\,\Delta_i P(\vJ,\vDelta)
\quad\hbox{and}\quad
\overline{\Delta^2_{ij}}(\vJ)\equiv\int\d^3\vDelta\,\Delta_i\Delta_j
P(\vJ,\vDelta).
\]
 Equation (\ref{eq:FP}) is the \deffn{orbit-averaged Fokker-Planck equation}.
It states that the rate of change of the distribution function is minus the
divergence of the flux $\vF$ of stars in action space, and we have an
expression for that flux in
terms of the \deffn{diffusion coefficients} defined by equations
(\ref{eq:diffcoeffs}). The latter are simply the expectation values and the
variances of the probability distribution of changes in actions per unit time.

\subsubsection{Diffusion coefficients}\label{sec:diffcoeffs}

The diffusion coefficients reflect the physics of whatever is responsible for
causing the fluctuations. In some circumstances, for example in a star
cluster, the fluctuations will be
approximately thermal in nature, with temperature $T$. Then the
principle of detailed balance requires that the stellar flux vanish
when the objects
being scattered are in thermal equilibrium with the fluctuations. That is,
$\vF=0$ for
 \[
f(\vJ)=\const\times\e^{-H/kT},
\]
 where $H(\vJ)$ is the Hamiltonian. When we substitute this form of the \df\
 into the equation $\vF=0$ and observe that for this \df
 \[
{\pa f\over\pa J_i}=-f{\Omega_i\over kT},
\]
 we obtain  an expression for the
first-order diffusion coefficient in terms of the second-order coefficient
\citep{BinneyL}
 \[\label{eq:BinneyLforT}
\overline{\Delta_i}=\fracj12\left({\pa\overline{\Delta^2_{ij}}\over\pa
J_j}-{\Omega_j\over
kT}\overline{\Delta^2_{ij}}\right).
\]
 This expression is useful because it enables us to obtain the
\deffn{first-order diffusion coefficients} $\overline{\Delta_i}$ from the
\deffn{second-order diffusion coefficients} $\overline{\Delta^2_{ij}}$, and,
while $\overline{\Delta^2_{ij}}$ can be obtained from first-order
perturbation theory (see below), a direct calculation of
$\overline{\Delta_i}$ requires second-order perturbation theory
\citep[see Appendix A of][]{BinneyL}.

We calculate the second-order diffusion coefficients by expanding the
potential in angle-action coordinates
 \[\label{eq:Phin}
\Phi(\vx,t)=\Phi_0(\vx)+\Phi_1(\vx,t)=\Phi_0
+\sum_\vn\Phi_\vn(\vJ,t)\cos(\vn\cdot\vtheta+\psi_\vn),
\]
 where $\Phi_0(\vx)$ is the potential of the underlying Hamiltonian
$H_0(\vJ)$ and $\Phi_1$ is the fluctuating part of the potential. By
integrating the equations of motion of the first-order change in
the actions for a time $T$ longer than the autocorrelation time of the
fluctuations we may show  \citep[see][for details]{BinneyL,Binney2012c} that
 \[\label{eq:Deltfromc}
\overline{\Delta^2_{ij}}=\fracj12\sum_\vn
n_in_j\widetilde c_\vn(\vJ,\vn\cdot\vOmega_0),
\] 
 where $\widetilde c_\vn(\vJ,\omega)$ is the power spectrum of the
fluctuations:
 \[
\widetilde c_\vn(\vJ,\omega)\equiv
\int_{-T}^T\d v\,\overline{\Phi_\vn(\vJ,t)\Phi_\vn(\vJ,t-v)}\cos(\omega v).
\]

The key implication of equation (\ref{eq:Deltfromc}) is that the ability of a
star to diffuse through phase space hinges on whether the fluctuations
contain power at one of the star's natural frequencies $\vn\cdot\vOmega_0$.
When the potential is constant in a steadily rotating frame, the expansion
coefficients $\Phi_\vn(\vJ,t)$ will contain only multiples of $\Omegap$ in
their temporal Fourier transforms. For example, if the potential is that of
an $m$-armed spiral, the potential will be $\propto\cos(m\Omegap t+\psi)$, so
the power spectrum of the potential $\widetilde c_\vn(\vJ,\omega)$ will be
non-zero only when $\omega=\vn\cdot\vOmega_0$ is equal to $m\Omegap$. In
other words, the only stars that will be caused to diffuse by an $m$-armed
spiral are those for which $\vn\cdot\vOmega_0=m\Omegap$ for some $\vn$
\citep{BarbanisW,LyndenBK}.  Besides the \CR\ [$\vn=(0,m,0)$], the two most
important resonances are the \deffn{inner Lindblad resonance} (\ILR), where
$\vn=(-1,m,0)$, and the \deffn{outer Lindblad resonance} (\OLR), where
$\vn=(1,m,0)$. At the \ILR\ the D\"oppler-shifted frequency at which a star
perceives the spiral is $m(\Omega-\Omegap)$ and this coincides with its
radial frequency, $\Omega_r$, while at the \OLR\ the perceived frequency of
the spiral is $m(\Omegap-\Omega)$ and this again coincides with $\Omega_r$.
We have shown that in a cold disc the spiral absorbs $L_z$ at the \ILR\ and
emits it at the \OLR, and that changes in $L_z$ heat the disc at both
locations.

Periodic fluctuations will depopulate narrow regions of phase space: stars
for which $\vn\cdot\vOmega_0$ is equal to any harmonic of the fluctuation
will be scattered to new actions and then cease to be resonant because
fundamental frequencies are functions of the actions.  \cite{SellwoodK} find
evidence for such action-space ``grooves'' in numerical simulations of
stellar discs and show that they can generate new spiral features, which in
their turn generate other grooves.

\subsubsection{Bar-driven migration?}\label{sec:barmigrate}

\cite{MinchevF10} have argued that bars drive radial migration much faster
than does the transient spiral structure as discussed in \S\ref{sec:migration}.  The main point \cite{SellwoodB} were making was
that resonant scattering at \CR\ changes $L_z$ without heating the
disc, so we cannot infer from the coolness of the thin disc near the Sun that
stars are still near their birth radii. The bar's \CR\ lies far interior to
the Sun, where the disc is probably not so cool, so heating-free migration
around the bar's \CR\ is not of observational interest, although it
undoubtedly takes place. The Sun lies near the bar's \OLR\ \citep{DehnenOp}
so changes in the angular momenta of local stars that are driven by the bar
{\it will\/} be accompanied by heating and {\it can\/} be constrained by the
coolness of the thin disc.

\cite{MinchevF10} rightly consider that the signature of scattering by the
bar is enhancement of the changes in $L_z$ around \CR\ and \OLR.  However,
they present evidence that the action of the bar is enhanced by transient
spiral structure, with the implication that the bar is not solely responsible
for changes in $L_z$. The role of spiral structure in the experiments
presented by \cite{MinchevF10} and \cite{Minchev+11} is unclear.  In
\cite{MinchevF10} the bar and spiral structure have fixed pattern speeds and
the disc consists of test particles. As we indicated above, a weak but steady
pattern will not engender long-term orbital diffusion: rather it will
restructure phase space as it is turned on and thereafter little will change.
Spiral structure at a different frequency surely has the capacity to prevent
the system settling down in the new order, by making the overall potential
non-periodic. This is in fact the physics of resonance overlap discussed by
\cite{Chirikov79}. However, in this case one might expect the frequency of
the spiral structure to play a bigger role than it seems to in the
experiments of \cite{MinchevF10}. 

\cite{Minchev+11} present experiments with self-consistent N-body models of
disc galaxies of low central concentration that appear to be strongly bar
unstable. There is clear evidence of strong orbital diffusion driven by the
bar, but no evidence that this diffusion is enhanced by spiral structure.
Indeed, in their model of an Sa galaxy, which shows the strongest diffusion,
the pattern speed of the bar decreases by $\sim20\%$ over the short lifetime
of the simulation, so the system is never in any danger of settling to a
steady configuration in phase space.

\subsubsection{Heating of the solar neighbourhood}\label{sec:heatsnhd}

It has been known since the 1950s that near the Sun younger stars tend to
have smaller random velocities than older stars.
\cite{SpitzerS} recognised that this phenomenon reflected \deffn{stochastic
acceleration} of disc stars by the Galaxy's fluctuating gravitational field.
The Hipparcos data permitted beautiful quantification of the phenomenon, and
it emerges that the velocity dispersion of a stellar cohort increases with
age as $\sigma\sim t^{0.35}$ \citep{AumerB}.

It's instructive to infer from this result how the diffusion coefficients
must scale with $|\vJ|$. We make two simplifying assumptions: (i) that the
dominant scatterers are much more massive than stars, and (ii) that the
velocity dispersions of groups of stars scale with the mean actions in the
group as 
 \[
\sigma_r\propto\sqrt{\langle J_r\rangle}\quad\hbox{and}\quad
\sigma_z\propto\sqrt{\langle J_z\rangle}.
\]
 These relations are exact in the epicycle approximation, in which the radial
and vertical oscillations of stars are harmonic, so for
example\footnote{Quite generally we have that $\Omega_rJ_r$ is equal to the
time-averaged value of $v_R^2$ along any orbit.} $J_r=E_R/\kappa$. Since
scattering must be dominated by giant molecular clouds and spiral arms, the
assumption of massive scatterers will be a good one. In thermal equilibrium
with such massive bodies, stars would have velocity dispersions that are
larger than those of the clouds and arms ($\sim7\kms$) by the square root of
the ratio of masses, so the stars' velocity dispersion would be $>1000\kms$.
Consequently, we can use equation (\ref{eq:BinneyLforT}) in the limit of
infinite temperature,\footnote{See Appendix B of \cite{BinneyL} for a
rigorous justification of this step.} when the action-space flux becomes
 \[\label{eq:simpflux}
F_i=-\fracj12\overline{\Delta^2_{ij}}{\pa f\over\pa J_j}
\]
 so the Fokker-Planck equation simplifies to
 \[\label{eq:BinneyL}
{\pa f\over\pa t}=\fracj12{\pa \over\pa J_i}\left(\overline{\Delta^2_{ij}}{\pa f\over\pa
J_j}\right).
\]
 
Stars are born on orbits that have non-negligible angular momenta $L_z\equiv
J_\phi$ but small values of $J_r$ and $J_z$. Consequently, a young population
is initially distributed in action space along the $L_z$ axis, and diffusion
of this population is predominantly away from this line, towards larger
values of $J_r$ and $J_z$. For this reason we neglect derivatives with
respect to $J_\phi$ in equation (\ref{eq:BinneyL}).

On the line $J_r=J_z=0$, the flux of stars $\vF$ (eq.~\ref{eq:simpflux}) must have
vanishing $r$ and $z$ components to prevent stars diffusing out of the
physical region $J_r\ge0$, $J_z\ge0$. As a consequence of this requirement,
in the vicinity of the $J_\phi$ axis the tensor $\overline{\vDelta^2}$ varies rapidly with
$J_r$ and $J_z$.  We neglect the much weaker dependence of $\overline{\vDelta^2}$ on
$J_\phi$ and consider $\overline{\vDelta^2}$ to be a function $\overline{\vDelta^2}(\vj)$ of the
two-vector
 \[
\vj\equiv(J_r,J_z).
\]

In problems involving the ordinary diffusion equation, a key solution is the
Green's function $\exp(-x^2/2t)/(2\pi t)^{1/2}$, which describes the spatial
distribution at time $t$ of particles injected at $x=0$ at time $t=0$.
Analogously, we seek a Green's function of
the form
 \[\label{eq:fsimilar}
f=t^{-2a}f_0(\vX)\quad\hbox{where}\quad
\vX\equiv {\vj\over t^a}.
\]
 In the solution (\ref{eq:fsimilar})
the mean value of $|\vj|$ will increase with time as $t^a$,
and the power of $t$ multiplying $f_0$ ensures that the total number of stars
$\int\d L_z\int\d^2\vj\,f$ is conserved as stars diffuse from the axis.  Suppose
$\overline{\Delta^2_{ij}}$ scales such that
$\overline{\Delta^2_{ij}}(k\vj)=k^b\overline{\Delta^2_{ij}}(\vj)$. Then
putting $k=t^{-a}$ we have
$\overline{\Delta^2_{ij}}(\vX)=t^{-ab}\overline{\Delta^2_{ij}}(\vj)$.
Evaluating both sides of equation (\ref{eq:BinneyL}) with these assumptions
yields
 \[
-{1\over t^{2a+1}}\left(2af_0+a\vX\cdot{\pa f_0\over\pa \vX}\right)=
\fracj12 t^{ab-4a}{\pa\over\pa X_i}\left(\overline{\Delta^2_{ij}}(\vX){\pa f_0\over\pa X_j}\right).
\]
 This equation can be valid at all times only if $2a+1=4a-ab$, so $b=2-1/a$.
Consequently, the empirical result $\langle J_r\rangle\sim\sigma_r^2\sim
t^{2/3}$ implies $a\simeq\fracj23$ and $b\simeq\fracj12$. 

The scaling $\sigma_r\sim t^{1/2}$, which has been advocated by \cite{Wielen}
and several subsequent authors, implies $a=b=1$. A simple argument shows that
it is implausible for the diffusion coefficients to grow so rapidly with
$|\vJ|$. In the epicycle approximation, $J_r$ differs from the epicycle
energy $E_R$ only by the (constant) epicycle frequency, so $\Delta_r\sim
\Delta E_R=\vv\cdot\delta\vv$, where $\delta\vv$ is the projection into the
equatorial plane of the change in a star's velocity as a result of a
scattering event. Hence $\langle\Delta_r^2\rangle\sim |\vj|$ implies
 \[
E_R\sim\langle(\Delta
E_R)^2\rangle\sim\langle (\vv\cdot\delta\vv)^2\rangle\sim\langle E_R|\delta
\vv|^2\rangle.
\] 
 That is, $\sigma_r\sim t^{1/2}$ implies that $|\delta\vv|$ is independent of
$|\vv|$. However, gravitational scattering always causes the momentum change
$\delta\vv$ to decrease with increasing speed because the gravitational force
is independent of speed and the time for which it acts decreases as
$1/|\vv|$. 

Can we derive $\overline{\Delta^2_{ij}}(k\vj)\sim
k^{1/2}\overline{\Delta^2_{ij}}(\vj)$ from physics? \cite{BinneyL} show that
this scaling {\it is\/} predicted by the model of cloud-star scattering that
was introduced by \cite{SpitzerS}. In this model the clouds move at speed
$v_c$ on circular orbits and the stars are on orbits of non-zero
eccentricity. The scattering of the star by the cloud is assumed to be
essentially complete in a distance that is small compared to the epicycle
radius -- thus the impact parameters of star-cloud collisions need to be
small compared to epicycle radius. With these restrictions, the scattering
can be analysed in the rotating frame in which the cloud is stationary. In
this frame the Jacobi constant $\fracj12v^2+\Phi_{\rm eff}$ is invariant, so
the speed $v$ with which the star recedes from the cloud is equal to its
approach speed: in this frame the cloud merely deflects the star. Such
deflections have two important effects.  First, they can transfer energy
between motion within the plane and motion perpendicular to the plane.
Second, by increasing $v_\phi$ at the expense of $v_R$, or vice versa, they
can change the magnitude of the epicycle energy $E_R$, which can be written
\citep[][\S3.2.3]{BT08}
 \[
E_R=\fracj12v_R^2+\fracj12\gamma^2(v_\phi-v_c)^2,
\]
 where $\gamma\equiv2\Omega_\phi/\kappa$ is the twice ratio of the circular
and epicycle frequencies. Since $\gamma^2>1$, a deflection that decreases
$v_R$ and increases $|v_\phi-v_c|$ increases $E_R$, and conversely for
deflections that increase $v_R$. It turns out that deflections that decrease
$v_R$ predominate, so overall deflections increase random velocity. It's
worth noting that the energy required to heat the disc comes not from the
clouds but from the ample store of energy in the rotation of the stellar
disc: the clouds catalyse an outward transfer of angular momentum by
absorbing $L_z$ from stars that have guiding centres interior to their orbits
and handing it out to stars with larger guiding centres.

Unfortunately, the demonstration by \cite{BinneyL} that the Spitzer-Schwarzschild
model implies that $\Delta_{ij}^2(\vj)\propto|\vj|^{1/2}$
is defective in two respects: (i) it assumes that the relative velocity
with which a star encounters a cloud is dominated by epicycle motion rather
than differential rotation, and, more seriously, (ii) it assumes that stars
are confined to the equatorial plane. In reality as a star ages it oscillates
with increasing amplitude and period perpendicular to the plane, and these
oscillations decrease its probability of being scattered by a cloud.
Consequently, when this effect is taken into account,
$\overline{\Delta^2_{ij}}(\vj)$ increases with $|\vj|$ more slowly than as
$|\vj|^{1/2}$.

\cite{BinneyL} show that three-dimensional scattering by molecular clouds
generates a  tensor of diffusion coefficients $\overline{\Delta^2_{ij}}$
which is highly anisotropic. The consequence of this anisotropy is that we
expect $\sigma_z/\sigma_r\sim0.8$, which is significantly larger than the
observed value, $\sim0.6$.  \cite{Sellwood} argues that the discrepancy
arises from the erroneous assumption of an isotropic distribution of
encounters: as in two-body scattering, distant encounters are important, and
since both stars and clouds lie within the disc, distant encounters are
dominated by the velocity components that lie within the plane and do not
change $J_z$.

Thus it seems that scattering of stars by giant molecular clouds may set the
ratio of the vertical and horizontal velocity dispersions of disc stars.
While star-cloud scattering makes a significant contribution to the secular
increase in the velocity dispersions of stars, it probably cannot account
fully for the data because its effectiveness declines rapidly with increasing
velocity dispersion and thus cannot account for the numbers of stars with
radial dispersions $\gta30\kms$.

\begin{figure}
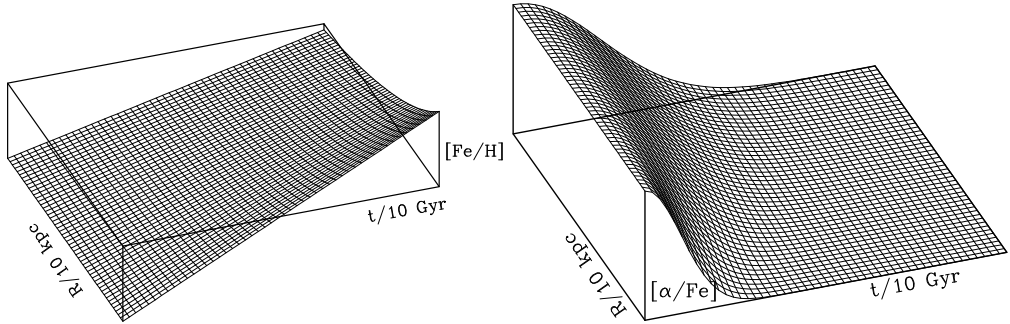

\centerline{\includegraphics[width=.48\hsize]{chemo1.ps}
\includegraphics[width=.48\hsize]{chemo0.ps}}
\caption{Sketches of plots of $\feh$ and $\afe$ as functions of time and
radius. Measured values of $\feh$ and $\afe$ each restrict the coordinates
$(R,t)$ of a star's birth to a curve: the curve in which the plotted surfaces
intersect the horizontal planes at the measured levels. The intersection of
these curves gives the  location.)}\label{fig:crossing}
\end{figure}

\section{Chemodynamical evolution}\label{sec:chemodyn}

It is more than half a century since it emerged from the work of
\cite{Roman50} and others that when stars are divided by chemistry, the
different groups have distinct kinematics. This phenomenon arises because
age and chemistry are correlated through the metal-enrichment history of the
ISM. 

Abundances correlate with ages in two principal ways. Stars formed in the
first $\sim\hbox{Gyr}$ have enhanced abundances of $\alpha$ elements (Mg, Ca,
Si, etc.) relative to iron because deflagration supernovae (\SNIa) are major
producers of Fe and, being the endpoints of binary-star evolution, they have
a gestation periods $\gta1\Gyr$ \citep[e.g.][]{Podsi}. In addition to a rapid
decline in $\afe$, we expect a longer-term increase in \feh. To the extent
that all the stars of the thin and thick discs formed near the plane, where
the star formation is certainly now concentrated, the number pair
$(\afe,\feh)$ should encode the radius and time of a star's formation because
plots of $\afe$ and $\feh$ as functions of $t$ will have different forms at
different radii (\figref{fig:crossing}), so from a star's value of $\afe$ we
can argue that its time and place of birth, $(t_{\rm b},R_{\rm b})$ lie on a
given curve in $(t,R)$ space.  The star's value of $\feh$ restricts $(t_{\rm
b},R_{\rm b})$ to a different curve, so $(t_{\rm b},R_{\rm b})$ can be
inferred from the intersection of these curves.

Unfortunately, this inference can be made only after the surfaces sketched in
\figref{fig:crossing} have been computed with a model of chemical evolution,
and, as we described in \S\ref{sec:simulations}, such models are subject to
many uncertainties. The general idea of a chemical-evolution model is that
stars form on near-circular orbits at rates determined by the local ISM
density. After birth the stars diffuse through action space.  The more
massive stars will at some point reach the ends of their lives and eject much
or all of their mass as metal-enriched gas, and perhaps leave a degenerate
remnant. A portion  of the gaseous ejecta will eventually join the cold ISM and
modify the chemical composition of subsequent generations of stars. To
compute such a model one first modifies the Fokker-Planck equation
(\ref{eq:FP}) to include a source of stars on circular orbits. That is, one
writes
 \[\label{eq:FPsource}
{\pa f\over\pa t}+{\pa F_i\over\pa J_i}=B(L_z,t)\delta(J_r)\delta(J_z),
\]
 where $B(L_z,t)$ specifies the rate at which stars form at the radius
$R_{\rm c}(L_z)$ of the circular obit of angular momentum $L_z$. This rate is
expected to be determined by the surface density of gas at radius $R$,
$\Sigma_{\rm g}(R,t)$. One needs to solve equation (\ref{eq:FPsource}) for a
\df\ $f(\vJ;\feh,\afe,\tau)$ that depends on chemical variables and age
$\tau$ in addition to $\vJ$, and from the developing solution and the theory
of stellar evolution determine the radial distribution of ejection of H, He,
Fe and $\alpha$ elements at each time and use these data to update the
chemistry of the ISM at each radius. 

As we discussed in \S\ref{sec:simulations}, it is incredibly hard to predict
which mixture of heavy elements will over time be ejected from a cohort of
stars because it hinges on details the life and death of massive single stars
and less massive binary stars that are extremely complex and delicate. This
fact is most unfortunate for these ``yields'' of heavy elements are key
ingredients of any model of chemodynamical evolution.  

Another important ingredient is the diffusion tensor $\overline{\vDelta^2}$
that determines the action-space flux (eq.~\ref{eq:simpflux}). The classical
work on chemical evolution \citep{PagelP75,Matteucci} restricted attention to
the solar cylinder, and later work \citep[e.g.][]{Chiappini03} neglected
diffusion in $L_z$. These restrictions greatly diminish the realism and
predictive power of the models.  \cite{SchoenrichB09a} extended the classical
work to include diffusion in orbit space, but did not use action integrals
and used a very crude scheme for computing changes in inclination.  This work
needs to be revisited using a more rigorous and transparent approach to
diffusion through action space.

The third and most uncertain ingredient is the dynamics of the ISM. A key
conclusion of the earliest studies of chemical evolution was that the
distribution in $\feh$ of local long-lived stars is inconsistent with the
solar neighbourhood having started with a fixed stock of gas that then turned
into stars as a closed system \citep{vdBergh62}. The data then available
could be explained either by a limited initial gas supply being constantly
topped up by fresh, metal-poor gas, or by the original gas being pre-enriched
by stars in the bulge and/or halo \citep{PagelP75,OstrikerT}.  An indication
that there has been accretion of metal-poor gas is that the star-formation
rate near the Sun seems to have fallen by only a factor $\sim3$ in the
$\sim10\Gyr$ life of the thin disc \citep{AumerB}, and given that the
star-formation rate must be a sensitive function of the density of the cold
ISM it implies that the mass of cold ISM has decreased by no more than a
factor 3 over that time. This being so, there simply wasn't enough mass in
the ISM $10\Gyr$ ago to form the present disc: much of the mass now in stars
must have been accreted. This line of argument is strongly supported by
measurements of the density of deuterium in the interstellar medium because,
although deuterium is rapidly destroyed within any star, the abundance in the
current ISM is lower than the primordial abundance by only a factor $\sim1.4$
\citep{Prodanovic}. This result clearly indicates that the ISM is constantly
accreting material with a near-primordial deuterium abundance. Long-term
accretion by halos is very much in line with the predictions of CDM
Cosmology, but there is some confusion whether accretion is predominantly
smooth or lumpy (i.e., associated with mergers). \cite{Huillier} find that 77\% of
accretion onto Milky-Way type halos is smooth and only 23\% lumpy.
A major issue is whether smooth accretion delivers baryons in a cold flow
($T\lta10^4\K$) or via a galaxy's virial-temperature corona.
\cite{Marasco12} suggest how gas may be accreted from the corona.

\cite{BilitewskiS} have investigated the distribution of accreted gas in
$L_z$, which must control the radial distribution of infall. The essential
point is that the gas cannot all be accreted at large radii, outside the
optical disc, but must accrete to radii comparable to those at which most
star formation takes place. 

Another key uncertainty for chemical evolution models is the extent to which
gas spirals inwards through the stellar disc. The spiralling of gas in
through most of the disc is an inevitable consequence of gas shocking
downstream of spiral arms, for then the surface density of gas has a phase
lag with respect to the density distribution of stars, and this phase lag
causes the gravitational field of the stars to drain angular momentum from
the gas. Most of the energy released by this accretion is radiated by the
post-shock gas.

When gas arrives just outside the corotation radius of the bulge/bar, it can
acquire angular momentum from the bar, so there is a tendency of gas to pile
up just outside the bar. Gas that crosses the bar's corotation radius loses
most of its angular momentum in a few dynamical times and is then dumped in
the Central Molecular Zone, a disc of radius $\sim200\pc$ made up of dense,
largely molecular gas on $x_2$ orbits \citep{Launhardt02,KimLee}.  This is a
region of intense star formation as is evidenced by the plethora of SN
remnants seen in radio continuum maps \citep{Kassim99} and the X-ray emitting
wind that emanates from this region \citep{CohenBH}.

One way of addressing these many uncertainties is to perform N-body plus
hydrodynamical simulations of the formation and chemical evolution of a
galaxy \citep[e.g.][]{Brook12,Loebman}. Ideally such simulations would be embedded in
a full cosmological context, but such simulations are still not capable of
resolving the detail that is crucial for studies of the solar neighbourhood
and the adjacent Galaxy (\S\ref{sec:simulations}).

\section{Spiral structure}\label{sec:spirals}

To this point we have largely ignored the self-consistency issue when
discussing perturbations: we have proceeded under the assumption that the
perturbing potential $\Phi_1$ is a given function of space-time. Ultimately
it is necessary to recognise that $\Phi_1$ is generated by the perturbation
$\rho_1$ in the mass density, but extending the mathematics to take this
constraint into account enormously increases the complexity of the problem.
In particular, it requires one to follow the dynamics of all the Galaxy's
many components, not just that of the particular stars that you are
observing.

The self-consistency constraint is of particular importance in connection
with the dynamics of the thin disc. Fortunately, this is the area in which
significant progress was made half a century ago through the work of Kalnajs,
Lin, Lynden-Bell, Toomre and their collaborators. From their work it emerged
that spiral disturbances, with their displacements either parallel to the
plane (forming spiral arms) or perpendicular to the plane (bending waves)
propagate radially. Bending waves simply propagate outwards with
ever-increasing amplitude as the disc's surface density declines. Eventually
their energy is converted into heat in the form of a flared outer disc
\citep{HunterT}. 

The situation regarding waves with displacements parallel to the plane is
much more complex \citep{Toomre81}.  These waves are excluded in a zone
around the \CR, and outside this exclusion zone, two wavenumbers are
permitted at a given frequency -- there are short- and long-wave branches of
the dispersion relation. Waves rattle between a Lindblad resonance and the
edge of the forbidden zone around \CR, steadily winding up as they travel: a
short-leading wave becomes first a long-leading wave, then a long-trailing
wave and finally a short-trailing wave before the wave finally thermalises
its energy at a Lindblad resonance. As the wave transitions from leading to
trailing form at a Lindblad resonance, the \deffn{swing amplifier} boosts its
amplitude by a factor that can be large. Since in a steady state the leading
waves in a disc have amplitudes that are smaller by this factor than those of
trailing waves, observed spirals have trailing morphology overall.  Moreover
solutions to the normal-mode equations have their largest amplitudes inside
\CR\ \citep{Toomre81}, and this is why spirals in the disc surrounding the
bar are constantly overtaken by the bar (\S\ref{sec:bulge}).

The results just described depend on the \deffn{tight-winding approximation}:
that the radial wavelength $\lambda_r\ll R$.  Unfortunately, this
is a poor  approximation on the long branches of the dispersion relation.
In view of this problem, the logical next step is the construction of global
normal modes by using angle-action coordinates to transform the normal-mode
problem into a problem in linear algebra. This has been done for a few
tractable discs \citep{Kalnajs77,ReadE} with the conclusion that the picture
of a normal mode as a superposition of trapped spiral waves is fundamentally
sound.

The pioneers of the theory of spiral structure hoped to be able to obtain a
complete understanding of the dynamics of discs in terms of waves and normal
modes \citep{LinShu}. This hope has since been dashed by numerical
experiments, which show that discs that have no unstable normal modes,
nonetheless demonstrate complex spiral instabilities that, given long enough,
grow from Poisson noise to O(1) amplitudes and ultimately lead to the
formation of a strong central bar \citep{Sellwood12}. The explanation of this
phenomenon is probably as follows. In the case of an electrostatic plasma it
is known that the set of ordinary normal modes is not complete
\citep{vanKampen}. That is, it
is not possible to express an arbitrary initial condition as a linear
combination of normal modes. So one cannot argue that a small initial
disturbance will stay small from the absence of growing modes. Gravitational
plasmas presumably share this incompleteness property, so a knowledge
of the stability of the ordinary normal modes can be used only to argue that
the system will be unstable if it has an unstable normal mode, and not that
it will be stable if all normal modes are stable.

The bottom line is that stellar discs are responsive dynamical systems
because they support waves that can be amplified by self gravity as they move
through the disc. The degree of amplification, and therefore the disc's
responsiveness, increases sharply as the velocity dispersion falls towards
the critical value at which Toomre's stability parameter \citep{Toomre64}
 \[
Q\equiv{\sigma_R\kappa\over3.36 G\Sigma}
\]
 falls to unity and the disc becomes unstable to radial
fragmentation.  Much of the energy carried by the waves is thermalised in the
vicinity of a Lindblad resonance. Thus the waves heat the disc and render it
less responsive.  

\subsection{Driving spiral structure}\label{sec:drivespirals}

By counting faint stars in the outer reaches of both our Galaxy and the
Andromeda nebula, M31, it has been shown that the outer parts of galaxies are
a mass of stellar streams and full of faint satellite galaxies
\citep{McConnachie,BelokurovFoS,Bell}. From studies of the internal dynamics
of satellite galaxies, we know that these systems are heavily dominated by
dark matter, so we must anticipate that the dark-matter distribution that
surrounds a galaxy like the Milky Way is lumpy. When a lump of dark matter
sweeps through pericentre, its tidal field will launch a wave into the host
galaxy's disc, which we know to be a responsive system. The classic example
of this process is M51, which has a satellite galaxy, NGC\,5195, near the end
of one of its exceptionally strong spiral arms. Few galaxies have such a
luminous satellite so near to them, so \deffn{grand-design} spirals like M51
are not prevalent.  Most galaxies will be responding simultaneously to more
than one much weaker stimulus, with the result that their spirals are both
weaker and rather chaotic.

\cite{MassetTagger} present convincing evidence that a bar excites a spiral
wave in the surrounding disc that has \ILR\ at the same radius as the bar's
\CR\ -- this is the real phenomenon of resonance overlap. Hence the Galactic
bar must be important for the Galaxy's spiral structure. In \S\ref{sec:bulge}
we saw that the the ``long thin bar'' inferred by \cite{Cabrera08} is nicely
explained by the more slowly-rotating spiral wave that the bar excites in the
surrounding disc. Also the ``molecular ring'' that dominates
the longitude-velocity plot of the Galaxy's CO emission \citep{DameHT} quite
likely consists of two giant spiral half-turns \citep{DobbsB} that the bar
generates in the gas disc.

\section{The warp}\label{sec:warp}

The early surveys of the Galaxy in neutral hydrogen revealed that the gas
disc is warped \citep{Burke57,Kerr57,Westerhout57}. Later it emerged that
the gas discs of external spiral galaxies are routinely warped \citep{Bosma,Briggs},
so the phenomenon must be a generic outcome of disc dynamics. Warps in
stellar discs are much harder to detect than warps in gas discs on account of
the difficulty of obtaining precise measurements of the stellar velocity
field, but it is clear that the Galaxy's stellar disc is warped
\citep{Dehnen98,Juric}. 

In the simplest approach to the dynamics of warps, we imagine that the disc
is made up of a series of rigid, spinning rings. The centre of each ring is
fixed, but can choose its own orientation in response to the torques exerted
by the ambient gravitational field. The latter is in part generated by the
rings themselves, and in part generated by the galaxy's halo and bulge. Each
ring intersects the galaxy's equatorial plane in two \deffn{nodes} on
opposite sides of the galaxy, and if the rings form a thin disc, the set of
all nodes traces a smooth curve through the galactic centre that is called
the \deffn{line of nodes}.  \cite{Briggs} reports that the line of nodes is
reasonably straight out to $\sim3R_\d$ and then becomes an open leading
spiral. A seductive picture of a warped galactic disc is one in which the
shape of the disc is always the same, but the line of nodes steadily rotates
because all the rings precess at the same rate. In the language of linear
analysis, the disc then has an excited $m=1$ normal mode.

It is helpful to consider the dynamics of the rings in two limiting
special cases. Consider first the case in which the rings are massless so the
only torques on them come from the gravitational field of the halo and bulge.
Then each ring precesses independently of the other rings at a frequency that
in general decreases outwards (being of order a fraction of the circular
frequency at the ring's radius) and increases with the flattening of the
gravitational field. Since $\Omega_z>\Omega_\phi$, the precession is in the
opposite sense to that in which stars circulate.
Unless the flattening increases with radius in a very
specific way, the precession frequency will vary with radius, with the
consequence that the line of nodes will \deffn{wind up} into an ever tighter
spiral. Hence a normal mode is not possible with massless rings.

Consider next the opposite case of very massive rings in which we can neglect
the contributions of bulge and halo to the gravitational field. Then the
rings form a system of coupled particles analogous to a chain of particles
that are linked by springs. Hence \deffn{bending waves} can propagate through
the disc, and we should ask whether normal modes can be constructed by waves
of a suitable frequency propagating from centre to the edge and then back,
returning to their point of departure with their original phase.  The answer
proves to be \citep{HunterT} that such standing-wave patterns are possible
only if the disc's outer edge is unrealistically sharp.  In the absence of a
sharp edge, the wavelength of an outwards propagating wave decreases, and its
amplitude grows, in an ever more painful attempt to transmit a constant flux
of mechanical energy outwards through a medium of ever decreasing density
(the whiplash effect).
When the wavelength become comparable with the typical amplitude of radial
oscillations of the disc's stars, the wave energy will be thermalised and
serve to increase the disc's thickness.  Thus the outer boundary of a disc
with a realistic density profile absorbs rather than reflects bending waves,
and normal modes cannot arise.

While these considerations show that a steadily precessing warped disc can
arise neither when the rings are very massive nor when they are very light,
for some years it seemed that steady precession would be possible in the
intermediate regime of moderately massive rings. To understand this case,
imagine ramping up the mass of a system of initially massless rings. Before
you ramp up the mass, the inner rings are precessing (backwards) faster than
the outer rings, and on account of the phase lag that develops between inner
and outer rings and an initially straight line of nodes will become a leading
spiral. When you make the mass non-zero, a torque will act between the inner
and outer rings. In the right circumstances this torque can slow the
precession of the inner rings and hasten that of the outer rings in just the
way required to set all rings precessing together with the line
of nodes remaining straight \citep{SparkeC} .

The problem with this model is that it treats the contribution of the bulge
and halo to the galactic potential as fixed. In reality the halo/bulge
experiences an equal and opposite force to that which it applies to the inner
rings, and it will respond dynamically to this force. N-body experiments show
that the response comes quite quickly, and tends to cause the bulge/halo to
align at each radius with the disc rings of comparable radius. Consequently,
if you set up a simulation of a system of rings embedded in an N-body halo in
what would be a normal mode if the bulge/halo were unresponsive, the
torques on rings from the bulge/halo quickly diverge from the values required
by the mode, and the line of nodes winds up on a dynamical timescale
\citep{BinneyJD}.

\cite{DubinskiC} assumed that the disc would be aligned with the shortest
principal axis of the inner halo (defined by $r<r_s$ or $r<2r_s$ where $r_s$
is the scale radius of the NFW halo) and determined the quadrupole component
of the (interior) gravitational potential due to the outer halo. They
analysed $\sim2000$ halos in this way to define the statistics of the
quadrupoles. Then they studied the response of an N-body disc to typical
quadrupoles. They found that the disc precessed and warped.  The precession
of the N-body discs could be largely reproduced using rigid, spinning discs,
so confirming that the self-gravity of the inner galaxy is strong enough to
render it equivalent to a rigid body right out to the edge, where warping
takes place.  The amplitudes of the warps and their morphologies were
entirely consistent with \cite{Briggs}. This work makes a very strong case
that warps are caused by misalignments between the inner and outer parts of
dark-matter halos. In doing so the study poses a serious problem for theories
of modified gravity, for in  these theories any quadrupole in the gravitational potential would be
generated by the inner galaxy (where all mass would reside) and would
consequently be aligned with the inner galaxy, so warps would not arise.

The misalignment of inner and outer halos is a natural consequence of
evolution of the pattern of accretion over cosmic time: the inner halo formed
long ago from an accretion pattern that reflected the structure of the cosmic
web at that time, while the outer halo is forming now and reflects the
current accretion pattern. Indeed, standard
cosmology predicts that the direction of angular
momentum vector of accreting material slews over cosmic time through at least
a radian \citep{QuinnB}. In exceptional circumstances this slewing can create
a disc in which roughly half the stars rotate one way and half the other, as
is observed in the lenticular galaxy NGC 4550 \citep{Rubinetal}. This outcome
is unusual, however, because when the angular momentum vector of accreting
gas slews gradually, by exchanging a small amount of angular momentum with
the stellar disc, gas will settle to the equatorial plane of the stellar disc
and corotating with the stars. In response to the exchange of angular momentum
with the gas, the spin axis of the stars shifts by a small amount in the
sense of reducing the offset between the disc's angular momentum vector and
that of the recently accreted gas. Through this mechanism we expect the
angular momentum vector of the stellar disc to remain roughly aligned with
that of currently infalling gas, so the latter always ends up corotating with
the stars. Nevertheless, freshly accreted material, baryonic and dark, will
have an angular momentum vector that is not parallel to that of the inner
Galaxy, and this offset naturally gives rise to a warp in the outer disc
\citep{JiangB}. In
fact gas discs within cosmological simulations of galaxies frequently have
warped outer zones.

\section{Outlook}\label{sec:outlook}

The study of our prototypical Galaxy is now recognised as a central task of
contemporary astrophysics. The challenge is to synthesise a huge body of
observational data from many different surveys at wavelengths from $21\,$cm
to X-rays and beyond into a coherent picture of gas, stars and dark matter.
Dynamical models are the key to this enterprise because they provide the
mechanism by which we can combine constraints from disparate surveys, and
map out the Galaxy's gravitational field. The last task is both crucial and
challenging: crucial because it provides our only realistic means to
determine the distribution of dark matter, which provides both the Galaxy's
backbone and our surest connection to cosmology, and challenging because the
dark-matter density is the divergence of the gravitational field, so we have
to determine the latter with sufficient precision to differentiate it. A
significant help in this enterprise should be the fact that the Galaxy
comprises many chemically distinct populations, each with its own
characteristic dynamics. Each  such population probes the common
gravitational field in a different way, so chemo-dynamical models should
prove powerful diagnostics of the dark-matter distribution.

The Galaxy is a complex machine, which inhabits a noisy environment. We need to
understand how it responds to that noise both because those responses will
impact the observational data, and because they contribute to driving the
Galaxy's evolution. The standard tool for computing the response of a
dynamical system is perturbation theory. To date perturbation theory has
played a rather modest role in Galactic dynamics, and if we are to make good
progress, this must change. In the eighteenth and nineteenth centuries the
best minds in mathematical physics struggled with the perturbative dynamics
of the solar system, and mastered it by inventing angle-action coordinates.
Since these coordinates provide the key to progress in Galactic dynamics,
they have been given prominence  in this article.

Our first task is to learn how the Galaxy is currently structured and
functions. Once that has been accomplished, it will be time to move on to the
still more challenging task of understanding how it arrived at its present
state from the cosmological initial conditions, which we believe we know.  At
this stage close interplay between perturbation theory and brute-force
simulations is likely to be very productive. It is unrealistic to hope that
we will ever have a cosmological simulation that yields a satisfactory
representation of the current Galaxy. The role of simulations is rather to
help us understand how different processes function and interact. If other
branches of physics are useful guides, we will obtain this understanding by
modelling simulations with perturbation theory, for perturbation theory has
historically been the language of physics through which we develop
understanding. In fact, perturbation theory will serve as intermediary
between our actual Galaxy and many different simulations, none of which is
quite the same as the Galaxy, but some of which resemble it in statistical
measures that will be defined by perturbation theory.

With the launch of the Gaia satellite in late 2013 and data from
complementary ground-based surveys such as the Gaia-ESO, APOGEE and Galah
surveys, our empirical knowledge of our Galaxy will take a quantum leap. Much
remains to be done before we can properly exploit the mines of data that
these surveys will produce.

\end{document}